\documentclass[aps,prx,twocolumn,showpacs,psfig,superscriptaddress,longbibliography]{revtex4-1}

\usepackage{times}
\usepackage{graphicx}
\usepackage{float}
\usepackage{latexsym,amsmath,amssymb,bm,euscript}
\usepackage{color}
\usepackage{subfigure}
\usepackage{epstopdf}
\usepackage[colorlinks=true,linkcolor=blue,citecolor=blue]{hyperref}
\usepackage{soul}
\usepackage[normalem]{ulem}
\usepackage{mathrsfs}
\usepackage{amsmath}
\usepackage{lettrine}
\usepackage{bbding}
\usepackage{xspace}
\usepackage{textcomp}
\usepackage{textcase}
\usepackage{setspace}

\newcommand {\B}{\textcolor {blue}}

\newcommand{\rev}[1]{{\color[rgb]{0,0,0}{#1}}}

\def\para{\ensuremath{/\kern -0.8em /}\xspace}
\def\ket#1{|{#1}\rangle}                 %% |#1>
\def\beqn{\begin{eqnarray}}
\def\eeqn{\end{eqnarray}}
\def\beq{\begin{equation}}
\def\eeq{\end{equation}}

\newcommand{\Beq}{\begin{eqnarray*} }
\newcommand{\Eeq}{\end{eqnarray*} }
\newcommand{\Bmat}{\left(\begin{matrix}}
\newcommand{\Emat}{\end{matrix}\right)}

\begin{document}

\title{Magnetocaloric Effect of Topological Excitations in Kitaev Magnets}

\author{Han Li}
\affiliation{Kavli Institute for Theoretical Sciences, University of Chinese 
Academy of Sciences, Beijing 100190, China}
\affiliation{CAS Key Laboratory of Theoretical Physics, Institute of Theoretical 
Physics, Chinese Academy of Sciences, Beijing 100190, China}

\author{Enze Lv}
\affiliation{CAS Key Laboratory of Theoretical Physics, Institute of Theoretical 
Physics, Chinese Academy of Sciences, Beijing 100190, China}

\author{Ning Xi}
\affiliation{CAS Key Laboratory of Theoretical Physics, Institute of Theoretical 
Physics, Chinese Academy of Sciences, Beijing 100190, China}

\author{Yuan Gao}
\affiliation{Peng Huanwu Collaborative Center for Research and Education, 
and School of Physics, Beihang University, Beijing 100191, China}
\affiliation{CAS Key Laboratory of Theoretical Physics, Institute of Theoretical 
Physics, Chinese Academy of Sciences, Beijing 100190, China}

\author{Yang Qi}
\affiliation{State Key Laboratory of Surface Physics and Department of Physics, 
Fudan University, Shanghai 200433, China}

\author{Wei Li}
\email{w.li@itp.ac.cn}
\affiliation{CAS Key Laboratory of Theoretical Physics, Institute of Theoretical 
Physics, Chinese Academy of Sciences, Beijing 100190, China}
\affiliation{Peng Huanwu Collaborative Center for Research and Education, 
and School of Physics, Beihang University, Beijing 100191, China}

\author{Gang Su}
\email{gsu@ucas.ac.cn}
\affiliation{Kavli Institute for Theoretical Sciences, University of Chinese 
Academy of Sciences, Beijing 100190, China}

%=====================================
\begin{abstract} 
\end{abstract}
\date{\today}
\maketitle

\noindent{\bf{Abstract}}\\
Traditional magnetic sub-Kelvin cooling relies on the nearly free local moments in hydrate 
paramagnetic salts, \rev{whose utility is hampered by the} dilute magnetic ions and low 
thermal conductivity. Here we propose to use instead fractional excitations inherent to 
quantum spin liquids (QSLs) \rev{as an alternative, which are sensitive to external fields 
and can induce a very distinctive magnetocaloric effect.} With state-of-the-art tensor-network 
approach, we compute low-temperature properties of Kitaev honeycomb model. \rev{For
the ferromagnetic case,} strong demagnetization cooling effect is observed due to the 
nearly free $Z_2$ vortices via spin fractionalization, \rev{described by a paramagnetic 
equation of state} with a renormalized Curie constant. For the antiferromagnetic Kitaev case, 
\rev{we uncover} an intermediate-field \rev{gapless QSL phase with very large spin entropy, 
possibly due to the emergence of spinon Fermi surface. 
Potential} realization of topological excitation cooling in Kitaev materials is also discussed, 
which may offer a promising pathway to circumvent existing limitations in the paramagnetic hydrates.
\\

\noindent{\bf{Introduction}}\\
The discovery of magnetocaloric effect (MCE) by Weiss and Piccard in 1917 
was a milestone in scientific discovery, bridging the disciplines of magnetics and 
calorics~\cite{Weiss1917,Smith2013}. Under the variation of magnetic fields, 
there occur a substantial entropy change and thus temperature variations under 
adiabatic conditions. In particular, sub-Kelvin cooling was achieved through 
adiabatic demagnetization refrigeration (ADR) with hydrate paramagnetic 
salts~\cite{Debye1926,Giauque1933}, which contain nearly free spins that exhibit 
prominent MCE. However, the paramagnetic coolants also suffer from intrinsic 
shortcomings, including low magnetic ion density, chemical instability due to the 
hydrate structure, and low thermal conductivity, etc. Currently, sub-Kelvin ADR 
plays an important role in space applications~\cite{Hagmann1995,Shirron2014}, 
and also holds great potential for helium-free cooling in advanced quantum 
technologies~\cite{Jahromi2019nasa}. Finding more capable magnetic materials 
for sub-Kelvin cooling is very demanding for addressing global scarcity of helium 
supply~\cite{Cho2009Science,Kramer2019Helium}. 

The low-dimensional quantum magnets have large ion density and stable 
structure, and may exhibit exotic spin states possessing high entropy density 
carried by the collective excitations. Cooling through many-body effects, 
they provide novel magnetocaloric materials and have raised great research 
interest recently~\cite{Zhu2003,Wolf2011,Lang2012,Tokiwa2016,Liu2021,
Liu2022,Xiang2024}. Typically, magnetic entropy gradually releases as spin 
correlations build up, and it becomes very small when certain spin ``solid'' 
order forms at sufficiently low temperature. To avoid such a classical fate, 
one could resort to highly frustrated magnets with strong spin fluctuations 
till low temperature. The quantum spin liquids (QSLs)~\cite{Anderson1973,
Balents2010,Zhou2017,JW2019QMats,Broholm2020Science}, resisting any 
magnetic ordering due to frustration effect and quantum fluctuations, present 
a particularly promising avenue for exploration~\cite{Liu2022}. \rev{Although
QSL systems hold significant potential, there is currently a gap in understanding 
how the unique properties of QSLs could be harnessed for advanced magnetic 
cooling.}

% ====================== Lattice ===================== %
\begin{figure*}[t!]
\includegraphics[angle=0,width=1\linewidth]{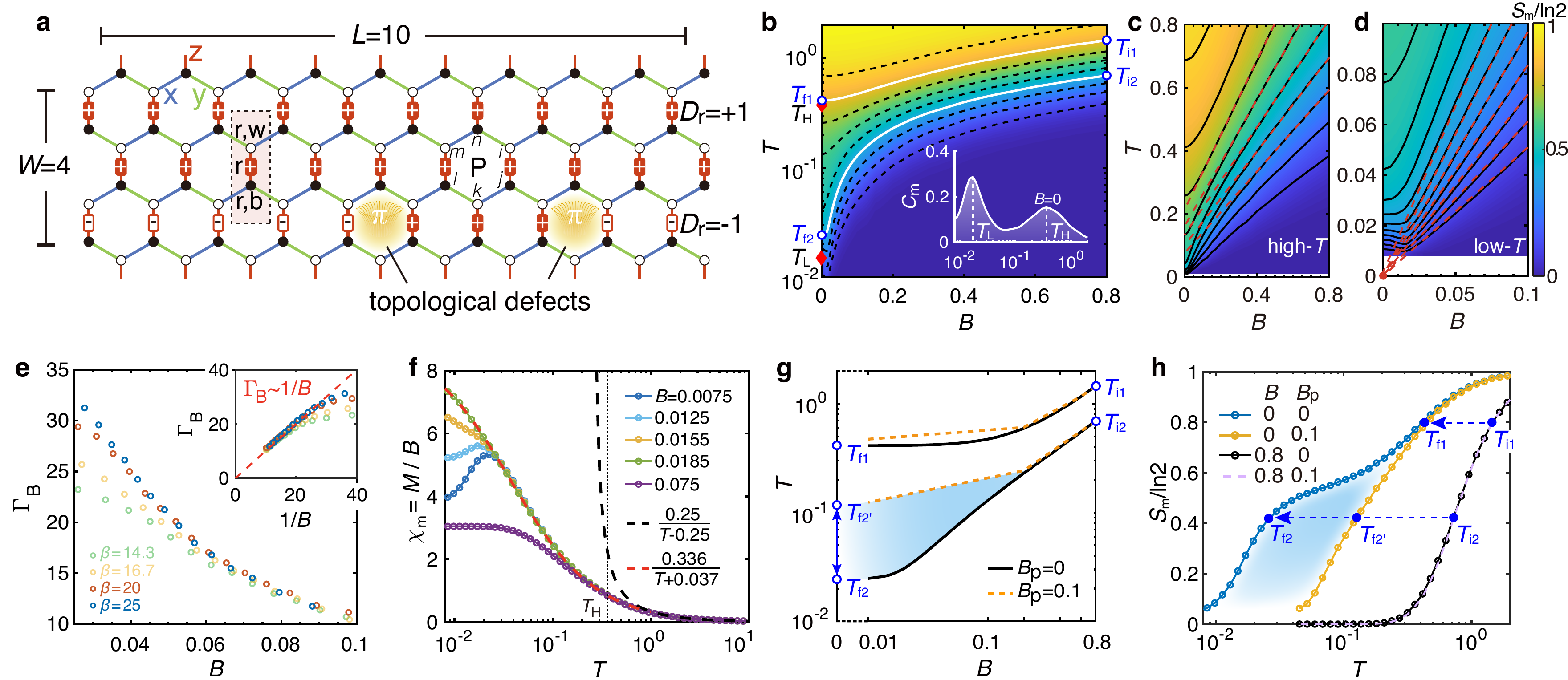}
\renewcommand{\figurename}{\textbf{Fig. }}
\caption{
\textbf{Kitaev paramagnetism and demagnetization cooling.} 
\textbf{a} Illustration of the Y-type cylindrical lattice and the topological excitations 
in the Kitaev model, where blue, green, and red bonds indicate respectively 
the $x$-, $y$-, and $z$-type interactions. The ``$+$'' (``$-$'') sign on the red 
bonds denote $D_{r}=+1$ ($-1$). 
A pair of $\pi$-fluxes (topological defects) can be created by
changing the sign of $D_{r}$ on a vertical bond (or an odd number of bonds).
\textbf{b} The landscape of isentropes for the FM Kitaev model with field $B$ up to 
$0.8$. At zero field, the specific heat $C_{\rm m}$ curve shows a double-peak 
feature at $T_{\rm L} \simeq 0.017$ and $T_{\rm H} \simeq 0.36$, as shown
in the inset. Two typical, and distinct ADR processes from the initial $T_{\rm i1(2)}$ 
to the final $T_{\rm f1(2)}$, are indicated with the white lines. \textbf{c} High-temperature 
isentropes following the Curie-Weiss behaviors and \textbf{d} low-temperature 
isentropes intersecting at the origin indicative of the emergent Curie 
paramagnetism. \textbf{e} The Gr\"uneisen parameter $\Gamma_{\rm B}$ at various 
low temperatures, which follows a $\Gamma_{\rm B} \sim 1/B$ behavior as shown 
in the inset. \textbf{f} The magnetic susceptibility $\chi_{\rm m}$ at various fields for the 
FM Kitaev model. The Curie-Weiss fitting at high ($T\gtrsim T_{\rm H}$) and 
Curie-law fitting at intermediate temperature ($T_{\rm L} \lesssim T \lesssim
T_{\rm H}$) are indicated by the black and red dashed curves, respectively. 
\textbf{g} The comparison of the ADR processes with and without the pinning field 
$B_{\rm P}=0.1$, and \textbf{h} shows the thermal entropy curves at field $B=0$ 
and $0.8$. Starting from $T_{\rm i2}$ at $B=0.8$, the temperature can be decreased 
to $T_{\rm f2}$ and $T_{\rm f2'}$ in the absence and under pinning field $B_{\rm P}=0.1$,
respectively. The former is clearly lower than the latter, as highlighted by the 
shaded regions in both \textbf{g} and \textbf{h}. {Source data are provided as a Source Data file.}
}
\label{Fig:Intro}
\end{figure*}
% ================================================== %

In this work, we study the MCE of QSLs in the Kitaev honeycomb system, 
employing exponential tensor renormalization group approach (Methods) 
\cite{Chen.b+:2017:SETTN,Chen2018,Lih2019,Li2022tan}. In the 
ferromagnetic (FM) Kitaev model, we discover a paramagnetic regime 
with nearly free $Z_2$ vortices, where the ADR isentropic lines follow 
a linear scaling with the constant ratio $T/B$. For the antiferromagnetic 
(AF) Kitaev case, we uncover a gapless QSL emerging at a remarkably 
low temperature scale, about 3\textperthousand~of the spin coupling 
strength, which gives rise to an even stronger cooling effect. 
Such a low temperature scale poses significant challenges for 
calculations, underscoring the remarkable nature of the gapless QSL.
The observed properties, including the specific heat, thermal entropy, 
spin-lattice relaxation rate, and spin structure factors, strongly suggest 
the presence of a gapless U(1) QSL with spinon Fermi surface.
Our findings establish a robust foundation for the development of 
magnetic cooling involving Kitaev QSLs and similar systems, which 
could be examined by conducting magnetocaloric experiments on 
candidate materials such as Na$_2$Co$_2$TeO$_6$.
\\

\noindent{\bf{Results}}\\
\textbf{The Kitaev model and spin fractionalization.} 
We consider the Kitaev honeycomb model under magnetic field $B$
applied along the [111] direction perpendicular to the honeycomb plane,
\begin{equation}
H= K \sum_{\langle i,j\rangle_{\gamma}} S_i^{\gamma}S_j^{\gamma}
-  {B} \sum_{i,\gamma} {S}_i^\gamma,
\label{Eq:model}
\end{equation}
where $K$ is the Kitaev interaction whose absolute value is set as 1
(energy scale), and $\langle i,j \rangle_{\gamma}$ with $\gamma = \{x,y,z\}$ 
represents the nearest-neighbor Ising couplings on the $\gamma$ bond 
as shown in Fig.~\ref{Fig:Intro}\textbf{a}. 

The Kitaev model has exactly solvable QSL ground states~\cite{Kitaev2006,
Hermanns2018}. At finite temperature, thermal fractionalization occurs 
(c.f., Supplementary Note \B{1}), with two types of excitations, namely, 
the Majorana fermions and $Z_2$ gauge fluxes, activated at very 
different temperature scales $T_{\rm H}$ and $T_{\rm L}$, respectively
\cite{Nasu2015,Motome2020,Li2020b}. Consequently, there 
exists a double-peak specific heat (c.f., the inset of Fig.~\ref{Fig:Intro}\textbf{b}) 
and quasi-plateau with fractional entropy ({$\frac{1}{2}\ln{2}$, see 
Fig.~\ref{Fig:Intro}\textbf{h}) between $T_{\rm H}$ and $T_{\rm L}$. 
We dub such an intermediate-temperature regime as Kitaev fractional liquid 
(KFL)~\cite{Nasu2015,Yoshitake2016,Motome2020,Li2020b} --- a correlated 
spin state that exhibits spin fractionalization. Intriguingly, although the Kitaev 
QSL may be fragile upon magnetic fields or other non-Kitaev interactions
\cite{Trebst2019,Patel2019,Motome2019}, the KFL regime at elevated
temperature is robust against these perturbations, \rev{different system 
sizes, and various magnetic fields directions~\cite{Motome2019,Li2020b}}. 

\textbf{Emergent Curie law and demagnetization cooling.} 
In Fig.~\ref{Fig:Intro}\textbf{b}, we show the thermal entropy $S_{\rm m}/\ln{2}$ computed
under magnetic fields $B$ up to $0.8 |K|$ for the FM ($K<0$) Kitaev model. The 
dashed lines represent the isentropes where ADR process follows: For initial 
temperatures $T_{\rm i} \gtrsim T_{\rm H}$, the isentropic lines are relatively flat, 
reflecting a weak field tunability of the correlated spins; however, when the initial 
temperature is below $T_{\rm H}$, the isentropes instead become very steep at 
small fields. Such a prominent cooling effect is rather unexpected for correlated 
spin systems, and we ascribe it to the {fractional excitations in the peculiar Kitaev systems}.

To be specific, at relatively high fields and temperatures, the $T$-$B$ isentropic 
lines follow an approximate linear behavior $T \propto B + \rm{const.}$ in 
Fig.~\ref{Fig:Intro}\textbf{c}, where the constant intercepts in the temperature 
axis reflect spin interactions in the Kitaev model. 
Nevertheless, in Fig.~\ref{Fig:Intro}\textbf{d}, we zoom in into the low-$T$ 
regime, $T\lesssim {0.1}$ and $B\lesssim {0.1}$, and find there is 
a linear scaling $T \propto B$ in isentropes that extrapolate to the origin, 
representing an emergent Curie-law paramagnetic behavior. The emergent 
paramagnetism can be further verified by computing the Gr\"uneisen parameter 
$\Gamma_{\rm B} \equiv 1/T(\partial T/\partial B)_{S_{\rm m}}$. At low temperatures, 
such as $T=0.05$ ($\beta=20$), we find a scaling $\Gamma_{\rm B} \sim 1/B$ 
as indicated in the inset of Fig.~\ref{Fig:Intro}\textbf{e}. Moreover, the 
magnetic susceptibility $\chi_{\rm m}$ is shown in Fig.~\ref{Fig:Intro}\textbf{f}, 
from which we find an emergent Curie-law behavior $\chi_{\rm m} \simeq 
\frac{C_{\rm K}}{(T+\theta)}$ with a renormalized Curie constant $C_{\rm K} 
\simeq 1/3$ and very small $\theta \simeq 0.037$ in KFL regime~\cite{Li2020b}. 
We emphasize that such $1/B$ scaling in $\Gamma_{\rm B}$ and Curie-law 
scaling in $\chi_{\rm m}$ for free spins now appears in the interacting spin system. 
It suggests the presence of nearly free degrees of freedom that carry significant 
spin entropies and appear as low-energy excitations in the Kitaev QSL system.
\\

\textbf{Equation of state in the Kitaev fractional liquid.} 
To understand the paramagnetic behaviors in the KFL regime, we drive 
equation of state to describe the gas-like, nearly free $Z_2$ vortices 
proliferated at finite temperature ($T > T_{\rm L}$). To start with, 
we apply an unitary Jordan-Wigner transformation of the Kitaev Hamiltonian
\cite{Feng2007},
\begin{equation}
\begin{split}
H=& \frac{iK_x}{4} \sum_{\langle r',w; r,b \rangle_x} \gamma_{r',w} \gamma_{r,b}
- \frac{iK_y}{4} \sum_{\langle r,b;r',w \rangle_y} \gamma_{r,b} \gamma_{r', w}\\
&- \frac{iK_z}{4} \sum_{r} D_{r} \gamma_{r,b} \gamma_{r, w},
\end{split}
\label{Eq:Majorana}
\end{equation}
where $\gamma_{r,b(w)}$ represents  the bond variable, and $D_{r} = 
i\bar{\gamma}_{r,b} \bar{\gamma}_{r,w}$ is related to the gauge flux 
$W_{\rm P} = D_{r}D_{r+1}$ on a hexagon containing vertical bonds 
$r$ and $r+1$ (c.f., Fig.~\ref{Fig:Intro}\textbf{a}), which is a $Z_2$ 
variable with values of $\pm 1$. The eigenstates of the Kitaev model 
can be labeled with these $Z_2$ variables on each hexagon, and in 
the ground state they take the same sign in the same row to ensure 
the absence of any $\pi$ flux ($W_{\rm P}=1$). 
Given one $D_{r}$ flipped, $\pi$ flux is introduced in two neighboring 
hexagons that have $W_{\rm P}=-1$. These $\pi$ fluxes can be regarded 
as topological defects, dubbed vison excitations in the $Z_2$ gauge 
field, that get activated \rev{near the low temperature scale $T_{\rm L}$ 
(c.f., Supplementary Note \B{1}) close to the flux gap~\cite{VisonGap2023}}.

The low-temperature ADR in KFL disappears once the $Z_2$ fluxes
are pinned. In Fig.~\ref{Fig:Intro}\textbf{g}, we introduce a pinning field 
$- B_{\rm P} \sum_{\rm P}{\sigma}_i^x {\sigma}_j^y {\sigma}_k^z 
{\sigma}_l^x {\sigma}_m^y {\sigma}_n^z$ coupled to the $Z_2$ 
fluxes and compare the ADR with and without $B_{\rm P}=0.1$, 
where $\sigma^{\gamma} = 2 S^\gamma$ is the $\gamma$-component 
of the Pauli matrix, and $\{i,j,k,l,m,n\}$ label the six sites in a hexagonal 
plaquette ``P''. From $B=0.8$ and 
$T_{\rm i2}\simeq 0.8$, the pure Kitaev model undergoes a dramatic 
temperature decrease to $T_{\rm f2}$ in the ADR process, while the 
cooling effect is much weaker when the pinning field is applied. This can 
be understood by checking the entropy curves in Fig.~\ref{Fig:Intro}\textbf{h}, 
where the pinning field can freeze the $Z_2$ flux and move the temperature
scale $T_{\rm L}$ towards higher temperature. Consequently, the quasi-plateau 
feature no longer appears under the pinning fields [see the yellow curve in 
Figs.~\ref{Fig:Intro}\textbf{g,h}]. 

As spin flipping in the Kitaev model can create a pair of visons, the latter is 
thus field tunable and intimately related with the emergent paramagnetism 
in KFL. A careful analysis (Methods) shows that here the \rev{emergent 
paramagnetic} state can be effectively described by the equation of state (EOS) 
$$M = C_{\rm K} B/T,$$ 
with $C_{\rm K} \equiv \sum_{j,\gamma}\langle S_{i_0}^{\gamma}S_j^{\gamma}
\rangle$ computed in the zero-field Kitaev model is the renormalized Curie 
constant. The EOS indicates that the induced magnetic moment is proportional 
to the field $B$ and inversely proportional to temperature $T$, which is the 
same as that of the ideal Curie paramagnet consisted of free spins. The only 
difference is the renormalized $C_{\rm K}$ that originates from the peculiar spin 
correlations in the Kitaev QSL.

% ================== XTRG Kitaev FL ================== %
\begin{figure}[t!]
\includegraphics[angle=0,width=1\linewidth]{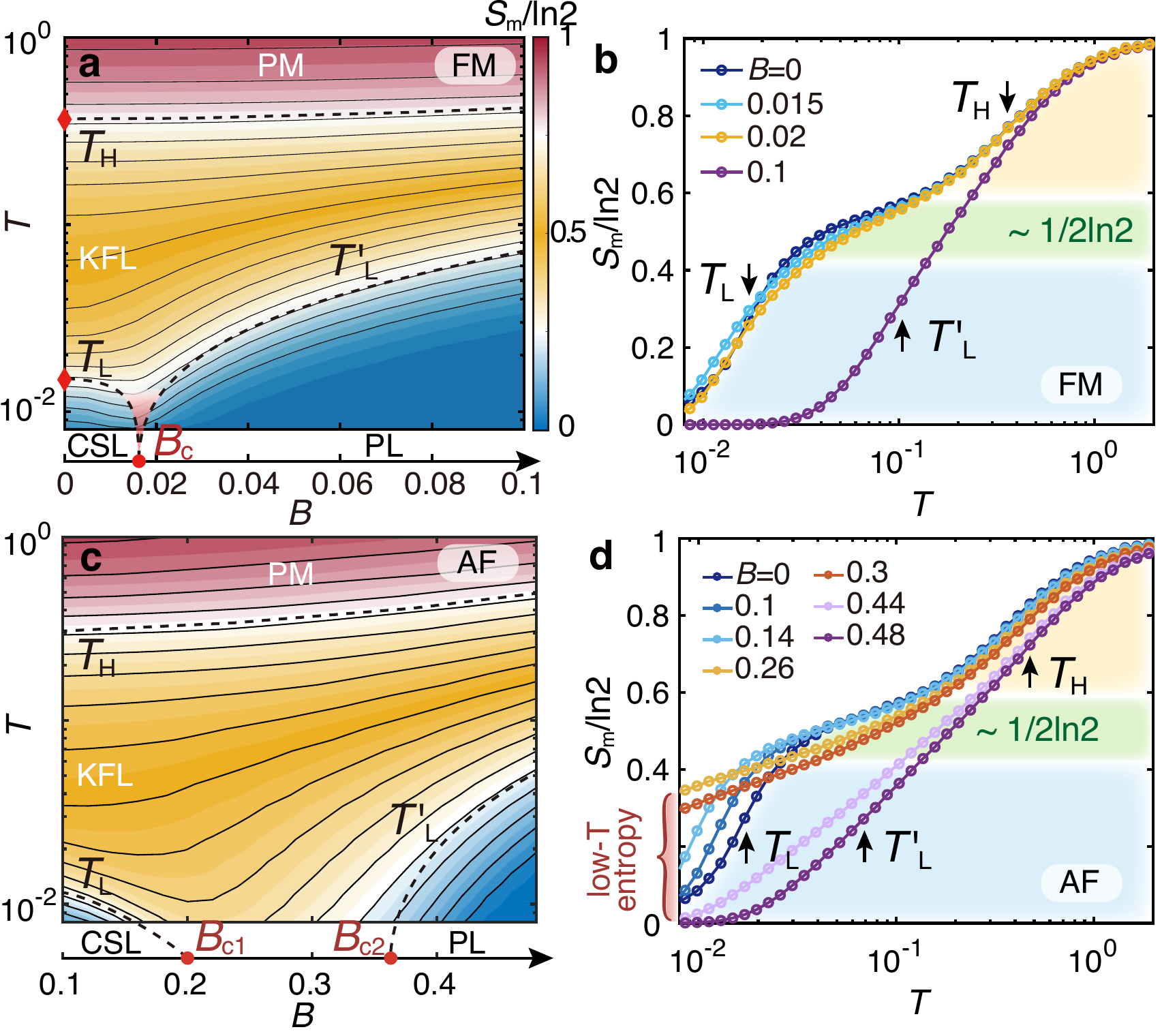}
\renewcommand{\figurename}{\textbf{Fig. }}
\caption{\textbf{The temperature-field phase diagrams and thermal entropy 
curves.} \textbf{a,c} The contour plots of thermal entropies and schematic 
temperature-field phase diagrams for the FM and AF Kitaev models 
at finite fields down to $T \simeq 0.008$. There are different regimes
in the phase diagram, i.e., the paramagnetic (PM), Kitaev fractional 
liquid (KFL), chiral spin liquid (CSL), and the polarized (PL) phase. 
The red dots on the horizontal axis denote the critical fields $B_{\rm c} 
\simeq 0.018$~\cite{Zhu2018} for FM and $B_{\rm c1} \simeq 0.2$ 
and $B_{\rm c2} \simeq 0.36$~\cite{Zhu2018,Patel2019} for the AF 
cases, as obtained with DMRG calculations. The shaded cone emerging 
from $B_{\rm c}$ in \textbf{a} indicates the quantum critical regime.
\textbf{b,d} The thermal entropy curves at various fields for the FM and 
AF Kitaev models, where the temperature scales $T_{\rm H}$, $T_{\rm L}$ 
and $T'_{\rm L}$ are indicated by the black arrows.
{Source data are provided as a Source Data file.}
}
\label{Fig:AFMIsentropes}
\end{figure}
% ================================================== %

% ==================  Gapless intermediate QSL ================== %
\begin{figure*}[t!]
\includegraphics[angle=0,width=1\linewidth]{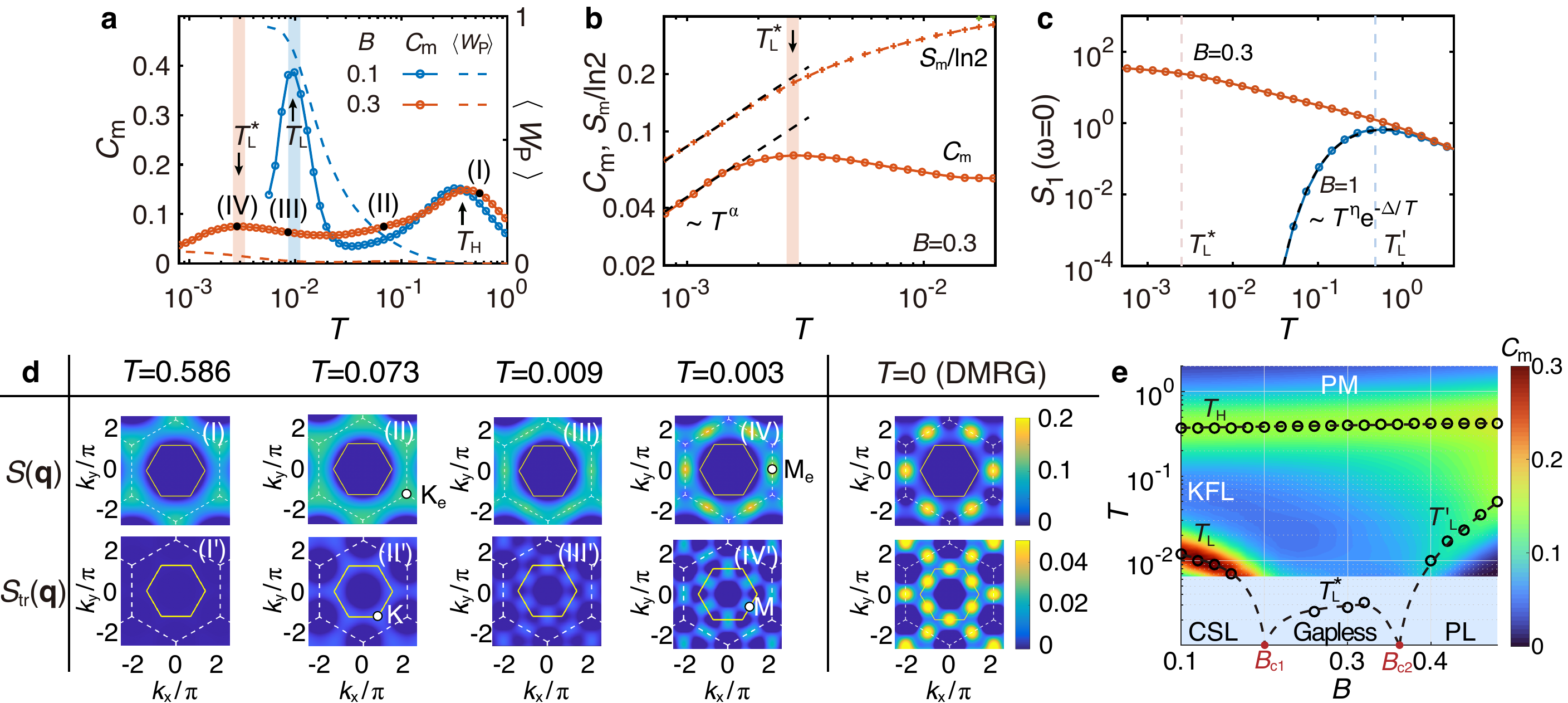}
\renewcommand{\figurename}{\textbf{Fig. }}
\caption{\textbf{The low-temperature calculations on the intermediate-field 
phase of the AF Kitaev model.} \textbf{a} The results of specific heat $C_{\rm m}$ 
and expectation $\langle W_{\rm P} \rangle$ computed on the YC$4 \times 
10 \times 2$ lattice under two typical fields $B=0.1$ and $0.3$. The calculations 
are performed down to an extraordinarily low temperature $T/K=8 \times 10^{-4}$. 
The temperature scales $T_{\rm H}$, $T_{\rm L}$, and the remarkably low 
$T^*_{\rm L}$ are all indicated by the arrows. \textbf{b} The log-log plot of the 
low-temperature $C_{\rm m}$ and $S_{\rm m}$/ln2 results under a field of $B=0.3$, 
where both curves show power-law scalings $T^{\alpha}$ \rev{($\alpha \simeq 0.8$)} 
below the low-temperature scale $T^*_{\rm L}$. \textbf{c} shows the estimate of 
relaxation rate $S_1(\omega=0)$ results at $B=0.3$ and $B=1$, where 
in the intermediate-field regime the calculated $S_1(\omega=0)$ continues 
to increase even below $T^*_{\rm L}$; while in the partially polarized phase 
it follows $T^{\eta} e^{-\Delta/T}$ ($\Delta \simeq 0.443$, $\eta \simeq -0.63$)
below $T'_{\rm L}$. 
\textbf{d} shows the temperature dependence of static spin structure 
factors $S({\textbf q})$ ($B=0.3$, see the main text) from (I) $T\simeq0.586$ 
to (IV) $T\simeq0.003$ (also marked in panel \textbf{a}). 
The corresponding $S_{tr}({\textbf q})$ for one sublattice are shown in the 
bottom panels (I$'$ to IV$'$). Representative high-symmetry points K$_{\rm e}$, 
K, M$_{\rm e}$ and M in the extended BZ are marked in (II),  (II$'$), (IV) and 
(IV$'$), respectively. The ground-state spin structure factor results obtained 
from DMRG are also displayed in \textbf{d}. \textbf{e} The contour plot of specific heat 
$C_{\rm m}$ down to $T \simeq 0.008$ with $B \in [0.1, 0.48]$. The black 
circles indicate the peak of the $C_{\rm m}$ curves, representing the 
temperature scales $T_{\rm H}$, $T_{\rm L}$, $T'_{\rm L}$ and $T^*_{\rm L}$ 
separating various magnetic phases, with the dashed line a guide for the eyes.
{Source data are provided as a Source Data file.}
}
\label{Fig:Gapless}
\end{figure*}
% ================================================== %

\textbf{Intermediate-field phase in AF Kitaev model.}
Beyond FM Kitaev model, we find such topological excitation MCE also 
in AF Kitaev system. As shown in Figs.~\ref{Fig:AFMIsentropes}\textbf{a,c},
the $B$ field applied along [111] direction can give rise to qualitatively 
different phase diagrams for the FM and AF isotropic Kitaev models
\cite{Gohlke2018,Zhu2018,Trebst2019,Jiang2011,Nasu2018}. For the 
FM case, we show magnetic entropy landscape with fields ranging from 
$B=0$ to $0.1$, where the dip of the isentropes gradually converges 
to the QCP $B_{\rm c} \simeq 0.018$~\cite{Jiang2011,Gohlke2018}. 

For the AF Kitaev model, on the other hand, we find two QCPs at 
$B_{\rm c1} \simeq 0.2$ and $B_{\rm c2}\simeq 0.36$ with an 
intermediate phase in between, whose nature is still under active 
investigations~\cite{Gohlke2018,Zhu2018,Liang2018,Jiang2018,
Trebst2019,Patel2019,Jiang2020}. In addition to magnetic entropy, 
the QCPs at $B_{\rm c1}$ and $B_{\rm c2}$ in the AF Kitaev model 
can also be identified through low-$T$ magnetization curves, matrix 
product operator entanglements, and spin-structure factors, etc., 
as shown in Supplementary Notes~\B{2,3}. 

The magnetic entropy curves vs. temperature are shown in Figs.
\ref{Fig:AFMIsentropes}\textbf{b,d}, where we compare the FM Kitaev 
model (Fig.~\ref{Fig:AFMIsentropes}\textbf{b}) with the AF case (Fig.
\ref{Fig:AFMIsentropes}\textbf{d}). In the former, we find the fractional
entropy remains robust in the KFL regime above the chiral spin liquid 
(CSL) phase (i.e., above the lower temperature scale $T_{\rm L}$). 
Such a quasi-plateau disappears for large field, like $B=0.1$, rendering 
a large entropy change driven by a relatively small field change.

Figure \ref{Fig:AFMIsentropes}\textbf{d} shows the magnetic entropy of 
the AF Kitaev model as a function of temperature for different magnetic 
fields. We observe that $T_{\rm L}$ shifts towards lower temperatures 
within the CSL phase, with the $\frac{1}{2}\ln{2}$ quasi-plateau feature 
remained. Moreover, in the intermediate-field regime, e.g., at $B=0.26$ 
and 0.3, the release of magnetic entropy is very slow, and $T_{\rm L}$ 
becomes no longer observable within the temperature window. 
As a result, a very prominent MCE occurs for the intermediate 
phase, \rev{which can be made more evident when employing units 
of measure such as Tesla for magnetic field and Kelvin for temperature
(see Supplementary Note~\B{4}). The lowest cooling temperature is 
found below 10~mK, given a proper Kitaev coupling strength and under 
a modest magnetic field change.} In the following, we exploit various 
finite-$T$ characterizations to clarify the nature of this intermediate-field 
phase and to understand the MCE in the AF Kitaev case. 

\textbf{Gapless QSL with possible spinon Fermi surface.} 
In Fig.~\ref{Fig:Gapless}\textbf{a}, we show the results of specific heat 
$C_{\rm m}$ and $Z_2$ flux $\langle W_{\rm P} \rangle$ for the AF 
case under out-of-plane fields. By pushing the calculations to an 
unprecedentedly low temperature $T/K\simeq 0.001$, we find a 
low-$T$ scale $T_{\rm L}^*\simeq 0.003$ indicated in Fig.
\ref{Fig:Gapless}\textbf{a} for the $B=0.3$ case, which is two orders 
of magnitude lower than $T_{\rm H} \simeq 0.3$. Considering 
the very small values of $\langle W_{\rm P} \rangle$ in Fig.
\ref{Fig:Gapless}\textbf{a}, we find $T_{\rm L}^*$ no longer reflects the 
$Z_2$ flux gap in the intermediate-field phase, but may be 
associated with other low-energy excitations.

% ================== XTRG Kitaev FL ================== %
\begin{figure*}[t!]
\includegraphics[angle=0,width=1\linewidth]{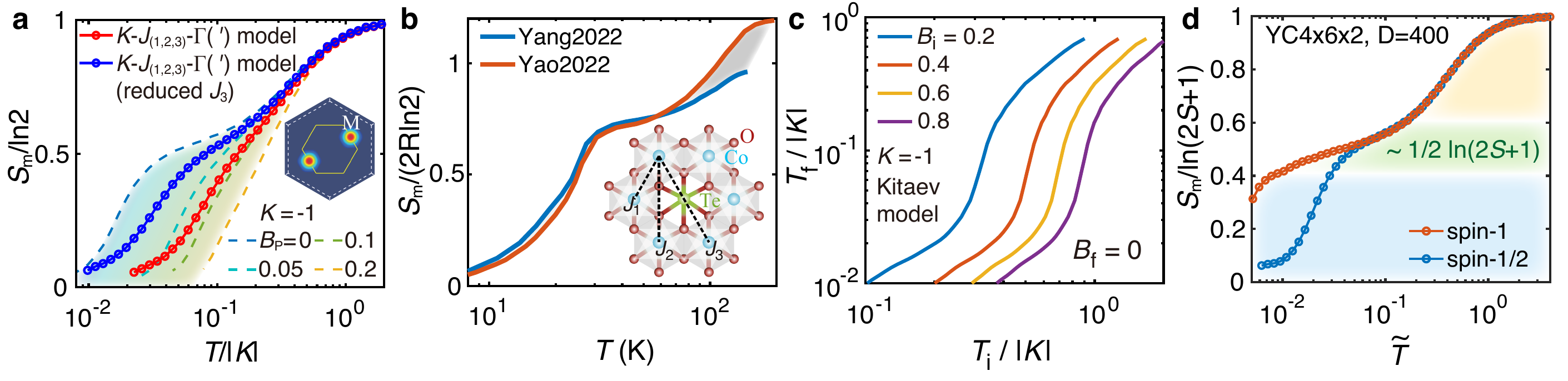}
\renewcommand{\figurename}{\textbf{Fig. }}
\caption{\textbf{Magnetic entropy analysis of Kitaev candidate 
material Na$_2$Co$_2$TeO$_6$ and higher-spin model.} \textbf{a} 
Simulated entropy data based on a realistic $K$-$J_{(1,2,3)}$-$\Gamma(')$ 
model~\cite{Samarakoon2021PRB}, the parameters (in natural unit) 
are $K=-1$, $J_1/|K|\simeq-0.03$, $J_2/|K|=0.007$, $J_3/|K|=0.17$, 
$\Gamma/|K|=0.003$, $\Gamma'/|K| = -0.033$. 
Although the quasi-plateau feature observed in the pure Kitaev model becomes 
blurred in the $K$-$J_{(1,2,3)}$-$\Gamma(')$ model (as shown by the red 
curve), a shoulder-like entropy can be discerned. When the strength of the 
$J_3$ is reduced by half, the low-$T$ entropy becomes greatly enhanced. 
\rev{The inset shows the zigzag AF order obtained with the realistic 
$K$-$J_{(1,2,3)}$-$\Gamma(')$ model in the ground state
\cite{Samarakoon2021PRB}, and} the shaded background with dashed 
lines represent the FM Kitaev model with various flux pinning fields 
$B_{\rm P}$. \textbf{b} The experimental results on Na$_2$Co$_2$TeO$_6$ 
measured by two groups~\cite{Yang2022,Yao2022PRL} with the shaded 
area highlighting the differences, and a schematic plot of crystal structure 
is shown in the inset. \textbf{c} The final temperature $T_{\rm f}$ reached 
at zero field as a function of the initial temperature $T_{\rm i}$ under various 
$B_{\rm i}$ from $0.2$ to $0.8$, providing an experimental test of Kitaev 
fractionalization in future MCE measurements. \textbf{d} The entropy curve
of spin-1 model compared to the spin-$1/2$ case, where the fractional
entropy plateau is more pronounced and extends to even lower temperature. 
The calculations are performed on YC$4\times6\times2$ systems with 
bond dimension $D=400$ in panel \textbf{d}. The rescaled temperature is
{$\widetilde{T} \equiv T/|K| \cdot \sqrt{{\rm ln}(2S+1)}/(S(S+1))$} (see Methods).
{Source data are provided as a Source Data file.}
}
\label{Fig:NCTO}
\end{figure*}

In Fig.~\ref{Fig:Gapless}\textbf{b}, we present the low-$T$ specific heat and 
entropy curves, which exhibit \rev{a power-law scaling $C_{\rm m} \sim T^{\alpha}$ 
with $\alpha \approx 0.8$ below $T_{\rm L}^*$.} This finding suggests a gapless 
nature of the intermediate-field QSL, and the \rev{sublinear power-law scaling 
in qualitative agreement with analytical results suggest the existence of a {U(1)} spinon 
Fermi surface~\cite{Trebst2019,Patel2019}. 
The emergence of U(1) gauge field and its coupling to spinons can significantly 
affect the low-energy properties~\cite{Shen2016}, leading to very soft modes 
and modified thermodynamic scalings with $\alpha < 1$~\cite{MotrunichPRB2005}. 
The results in Fig.~\ref{Fig:Gapless}\textbf{b} indicate a divergent $C_{\rm m}/T$, 
together with the observation of a specific heat peak at $T_{\rm L}^* \sim 0.003$,
indicating strong spinon-gauge fluctuations. They possibly account for the 
large spin entropy and explain the prominent MCE observed in Figs.~\ref{Fig:AFMIsentropes}\textbf{c,d}}.

In Fig.~\ref{Fig:Gapless}\textbf{c}, we show the spin-lattice 
relaxation rate $S_1(\omega=0)$ computed via an imaginary-time 
proxy~\cite{Xi2023}:
\begin{equation}
S_1(\omega=0) \equiv \frac{1}{T}\sum_{\gamma}\sum_{j=1}^{N}[\langle 
S_j^{\gamma}(\frac{\beta}{2}) S_j^{\gamma}(0)\rangle - \langle 
S_j^{\gamma}(\beta) \rangle^2],
\end{equation}
which probes the low-energy dynamics. In Fig.~\ref{Fig:Gapless}\textbf{c}, 
we observe that $S_1(\omega=0)$ continues to increase even below 
$T^*_{\rm L}$ for $B=0.3$, which indicates the \rev{strong spin fluctuations
and} gapless nature of the intermediate phase. Distinctly, $S_1(\omega=0)$ 
decays exponentially as $T^\eta e^{-\Delta/T}$ for $B=1$ in the gapped 
(partially) polarized phase.

To further explore the temperature evolution of the spin states, 
we show in Fig.~\ref{Fig:Gapless}\textbf{d} the spin structure factors 
${S}(\textbf{q})=\sum_{j\in {N}} 
e^{ i \textbf{q} (\textbf{r}_j-\textbf{r}_{i_0})} 
(\langle S_{i_0} S_j\rangle - \langle 
S_{i_0}\rangle\langle S_{j}\rangle)$,
where $i_0$ represents a central reference site, and the results are 
obtained by considering all sites and symmetrized over the ${\textbf q}$ 
points. When $j$ is restricted within one sublattice of the triangular 
lattice, we obtain a sublattice spin structure factor as $S_{\rm tr}({\bf q})$. 
In Fig.~\ref{Fig:Gapless}\textbf{d}, we show the calculated results of 
$S(\textbf{q})$ and  $S_{\rm tr}({\bf q})$ at various temperatures, 
where the structure factor peaks move from K$_{\rm e}$- to 
M$_{\rm e}$-point in the extended Brillouin zone (BZ) as the system 
cools down. It is noteworthy that there are still significant changes in
the spin structures even at very low temperature, which converge towards
the ground-state results only below $T^*_{\rm L}$ (c.f., the panels on 
the right column of Fig.~\ref{Fig:Gapless}\textbf{d}).

Based on the DMRG results of spin structure factor, a spinon-Fermi-surface 
U(1) QSL has been proposed, {with Fermi} pockets around the $\Gamma$ 
and M points in the real Brillouin zone~\cite{Patel2019}. The scattering 
function is constructed as $\sum_{\textbf q} \delta(\epsilon^S_F({\textbf q})) 
\delta(\epsilon^S_F({\textbf q}+{\textbf k}))$, \rev{where $\epsilon^S_{F}({\textbf q}) 
\equiv \epsilon({\textbf q}) - \epsilon_{F}$} and $\textbf{k}$ is the momentum 
transfer across the Fermi surface. Such spinon Fermi surface gives rise to a 
sublattice spin structure $S_{\rm tr}(\textbf q)$ with large intensity at the M 
points. \rev{As shown in the bottom panels in Fig.~\ref{Fig:Gapless}\textbf{d},
$S_{\rm tr}(\textbf q)$ develops M-peaks at temperature {around} $T^*_{\rm L}$, 
reaching a ``handshake'' with the DMRG calculations.}

Overall, our finite-$T$ results support the scenario of a 
gapless QSL, and the temperature-field phase diagram is shown 
in Fig.~\ref{Fig:Gapless}\textbf{e}. \rev{In the phase diagram, 
the high-temperature scale $T_{\rm H}$ determined by the 
spinon bandwidth is very robust and barely changes at different 
fields when changing from CSL to gapless U(1) QSL.}
It is worth noting that the energy scale {$T^*_{\rm L}$} is very small for 
the emergent gauge field in the intermediate-field phase, which requires 
high-resolution calculations to resolve its true ground state. 
This may explain the different conclusions obtained using various 
theoretical approaches and approximations, as discussed in previous 
ground-state studies~\cite{Patel2019,Trebst2019,Jiang2020,Zhang2022}. 

{\bf{Connections to realistic honeycomb-lattice magnets.}}
The Kitaev model can find its materialization in honeycomb-lattice magnets 
with significant spin-orbit couplings~\cite{Jackeli2009}. For example, the 
4$d$- and 5$d$-electron transition metal based compounds X$_2$IrO$_3$ 
(X = Na, Li, Cu)~\cite{Chaloupka2010,Singh2012,Yamaji2014,Choi2019} 
and XR$_3$ (X = Ru, Yb, Cr; R = Cl, I, Br)~\cite{McGuire2015,HSKim2015,
Kim2016,Ran2017,Winter2017NC,Wang2017,Banerjee2016,Do2017,
Banerjee2017,Imai2021,Hao2021}; the recently proposed 3$d$-electron 
Co-based honeycomb magnets~\cite{Liu2020PRL,Lin2021NC,Yao2022PRL,
Li2022,Zhong2020SA,Zhang2023NatMat,Halloran2023PNAS}; the rare-earth 
chalcohalide REChX (RE = rare earth; Ch = O, S, Se, Te; X = F, Cl, Br, I)
\cite{Zhang2021CPL} and Ba$_9$RE$_2$(SiO$_4$)$_6$ (RE = Ho-Yb)
\cite{Tian2023Honeycomb}; spin-1 honeycomb-lattice magnet 
Na$_3$Ni$_2$BiO$_6$~\cite{Shangguan2023} and spin-$3/2$ CrSiTe$_3$~\cite{XuPRL2020}, etc., have been proposed to accommodate 
Kitaev interactions. Although most of these compounds exhibit long-range 
magnetic order at sufficiently low temperature, signatures of Kitaev 
interaction and spin fractionalization~\cite{Do2017,Banerjee2018} 
have been observed in some of them.

Amongst others, the Co-based Kitaev magnet Na$_2$Co$_2$TeO$_6$ 
has recently attracted great research interest~\cite{Lin2021NC,Yao2022PRL,
Yang2022,Songvilay2020,Kim2021,Samarakoon2021PRB,LinG2022}. 
In Fig.~\ref{Fig:NCTO}\textbf{a}, we calculate the thermal entropy curves 
based on an effective $K$-$J_{(1,2,3)}$-$\Gamma(')$ model proposed 
in Ref.~\cite{Samarakoon2021PRB}, and compare them to experimental results 
in Fig.~\ref{Fig:NCTO}\textbf{b}.
{We note that} there are a number of extended 
Kitaev models~\cite{Songvilay2020,Kim2021,Lin2021NC,
Samarakoon2021PRB,LinG2022} with different parameter 
sets proposed for Na$_2$Co$_2$TeO$_6$, which share some similarities with 
the $K$-$J_{(1,2,3)}$-$\Gamma(')$ model adopted here. 
{Due to the presence of $J_{(1,2,3)}$ and $\Gamma(')$ terms, the ground 
state \rev{has a zigzag AF order (see inset of Fig.~\ref{Fig:NCTO}\textbf{a})} 
and deviates from a Kitaev QSL, {while} the \rev{magnetic entropy curve shows a 
shoulder-like feature.} This resembles the behavior observed in a pure FM Kitaev 
model with a pinning field $B_{\rm P}=0.1$ shown in Fig.~\ref{Fig:NCTO}\textbf{a}. 
We also compute the thermal entropy of a $K$-$J_{(1,2,3)}$-$\Gamma(')$ 
model with reduced $J_3$ term, \rev{where we observe a clearer signature 
of thermal fractionalization.} 
In Fig.~\ref{Fig:NCTO}\textbf{b}, the experimental data of magnetic 
entropy are plotted, which exhibit distinct plateau features and suggest 
a promising cooling capacity. However, there are differences observed 
between the two experimental curves from different groups~\cite{Yang2022,
Yao2022PRL}, possibly due to sample dependence, measurement errors, 
the way to dissociate the phononic and magnetic contributions, or possible 
electronic excitations beyond the $J_{\rm eff} =1/2$ manifold. 

\rev{Given the significant Kitaev interaction present in the effective model
considered, the emergence of a shoulder-like feature in our theoretical 
calculations --- a pattern mirrored in experiments on Na$_2$Co$_2$TeO$_6$
 --- suggests that we might be witnessing signatures of fractionalization 
phenomena, a hallmark of quantum entanglement.} We 
argue that the non-Kitaev terms in realistic compounds provide an effective} 
``pinning'' field $B_{\rm P}$,
which reduces the low-temperature entropy of topological excitations.
Additionally, there are also discrepancies between the simulated curves 
and the experimental ones, highlighting the urgent need to determine 
the \rev{precise microscopic spin model for} Na$_2$Co$_2$TeO$_6$. 

The results presented in Figs.~\ref{Fig:NCTO}\textbf{a,b} further demonstrate 
the robustness of spin fractionalization under moderate non-Kitaev interactions. 
Conversely, these results suggest that the emergence of paramagnetic 
behaviors could be an indicator of the presence of Kitaev interactions in 
realistic materials. {As shown in Fig.~\ref{Fig:NCTO}\textbf{c},} in practical 
ADR measurements one can decrease magnetic fields from various initial 
$B_{\rm i}$ to final $B_{\rm f}=0$ and measure the final cooling temperature 
$T_{\rm f}$. Besides Na$_2$Co$_2$TeO$_6$, {its} sister material 
Na$_3$Co$_2$SbO$_6$ has also been put forward to host the Kitaev 
interactions~\cite{Liu2020PRL}. Moreover, different from these two compounds 
having strong spin couplings comparable to $\alpha$-RuCl$_3$~\cite{Han2021}, 
we notice there are recent progresses in rare-earth honeycomb-lattice magnet 
Ba$_9$RE$_2$(SiO$_4$)$_6$ (RE = Ho–Yb)~\cite{Tian2023Honeycomb}
that have moderate couplings suitable for sub-Kelvin cooling. Our studies call 
for magnetic specific heat and in particular the MCE measurements on these 
honeycomb-lattice quantum magnets, \rev{which may provides a useful means 
to probe the Kitaev coupling.}
\\

\noindent{\bf{Discussion}}\\
To conclude, with the cutting-edge exponential tensor renormalization group 
approach~\cite{Chen2018} applied to the Kitaev systems, 
we construct comprehensive temperature-field phase diagrams for both $K<0$ 
and $K>0$ Kitaev models, where a linear $T$-$B$ curve in the ADR process is 
observed in the Kitaev fractional liquid regime. Moreover, for the AF case with 
$K>0$, we find thermodynamics evidence for intermediate-field gapless QSL 
with possible spinon Fermi surface and very pronounced magnetocaloric responses.

With this, we propose that Kitaev magnets hold not only potential applications 
in topological quantum computing but also in low-temperature refrigeration. 
Here, beyond the general argument of frustration effects, we establish a concrete 
connection between QSL physics and MCE through high-precision many-body 
calculations. \rev{The exotic fractional and topological excitations that are highly field-tunable open up new avenues for advanced magnetocalorics.}

On the other hand, unlike paramagnetic salts with nearly free local moments, 
here we reveal a significant cooling effect of the nearly free $Z_2$ fluxes 
arising from interacting spins. There are clear advantages of QSL coolants 
over paramagnetic salts. The ion density of the former can be one order 
of magnitude greater, and it thus renders much larger entropy density. 
Additionally, the hydrate paramagnetic salts suffer from low thermal conductivity 
and long relaxation time as the spins are diluted and isolated. On the contrary, 
in Kitaev QSL the spins fractionalize into localized fluxes and itinerant Majorana 
fermions. The latter exhibits metallic behavior and can enhance the thermal 
conductivity, making the Kitaev magnets truly exceptional candidates as 
helium-free quantum material coolants. Moreover, such a topological cooling also 
exists in higher-spin Kitaev systems, as shown in Fig.~\ref{Fig:NCTO}\textbf{d} 
(see also Methods), rendering a scalable cooling capacity with higher spins. 

Much like the exploration of low-temperature magnetocalorics on the 
triangular-lattice quantum antiferromagnet Na$_2$BaCo(PO$_4$)$_2$
\cite{Zhong2020SA,Gao2022} has expanded our knowledge with 
triangular-lattice spin supersolid and its giant cooling effect~\cite{Xiang2024}, 
we expect that the current proposal will lead to future discoveries and 
advancements in the studies of Kitaev materials. This represents a compelling 
approach to realize helium-free cooling by tapping into the topological excitations 
of emergent gauge fields within QSL systems and candidate materials.
\\

\noindent{\bf{Methods}}\\
\textbf{Density matrix and tensor renormalization group approaches.}
The ground state properties are computed by the density matrix 
renormalization group (DMRG) method, and the finite-temperature 
properties are computed with exponential tensor renormalization 
group (XTRG)~\cite{Chen2018,Lih2019}. As discussed in the main text, 
the two characteristic temperature scales in the original Kitaev model, 
i.e., $T_{\rm H} \simeq 0.36$ and $T_{\rm L}\simeq 0.017$ are 
separated by more than one order of magnitude. Therefore, 
it requires accurate and efficient many-body methods to carry 
out the low-temperature simulations under zero and finite magnetic fields. 

The XTRG method starts from a high-temperature density matrix 
$\rho_0(\tau_0) = e^{-\tau_0 H}$ with $\tau_0 = 0.0025$, whose matrix 
product operator (MPO) representation can be obtained accurately up 
to machine precision~\cite{Chen.b+:2017:SETTN}. By multiplying the 
MPO by itself, the system can be cooled down exponentially fast through 
$\rho_n \equiv \rho_{n-1} \cdot \rho_{n-1} =  \rho(2^n\tau_0)$, and the 
thermodynamic quantities like free energy, thermal entropy, specific heat,
as well as spin correlations, etc, could be calculated with high precision.
Such method has been employed to various 2D spin systems
\cite{Chen2018,Lih2019,Li2020b,Li2020,Han2021,Wang2023PRL,
Xiang2024}, which has been shown to be a highly efficient and powerful tool.
In DMRG, we keep up to $D=1024$ states that leads to a rather small 
truncation error $\epsilon \lesssim 1\times 10^{-8}$. In XTRG calculations,
with retained bond dimension $D$ up to ${600}$, we ensure the truncation 
error about $10^{-3} \sim 10^{-4}$ down to $T/|K| \simeq 0.001$ which well 
converge the results (c.f., Supplementary Note \B{2}). In the simulations, 
we mainly work with a Y-type cylindrical (YC) lattice YC$W\times L\times 2$ 
with width $W=4$ and length $L=10$, as illustrated in Fig.~\ref{Fig:Intro}\textbf{a}.
\\

\noindent{\textbf{High-spin Kitaev systems.} In Fig.~\ref{Fig:NCTO}\textbf{d} 
we show the entropy curve for the Kitaev model with higher spin $S=1$, 
as compared with the $S=1/2$ case. We find an even more prominent  
plateau-like structure with about $\frac{1}{2} \ln(2S+1)$ entropy. 
For general spin-$S$ Kitaev model, we consider a high-temperature 
expansion of the partition function up to the second order as
$Z(\beta) =  \left(2S+1\right)^N - \beta \mathrm{Tr}\left[H\right]
+ \frac{1}{2}\beta^2\mathrm{Tr}\left[H^2\right] + \mathcal{O}\left(\beta^3\right)$,
where $\mathrm{Tr}\left[H\right]=0$ and $\mathrm{Tr}\left[H^2\right]=\frac{1}{9} K^2S^2(S+1)^2$. As the high-temperature entropy reads 
$S_{\rm m}/N = \ln\left(2S+1\right) - \frac{1}{18}K^2S^2(S+1)^2/T^2$,
we can rescale the temperature as $\widetilde{T}
\equiv T/|K| \cdot \sqrt{{\rm ln}(2S+1)}/(S(S+1))$ to collapse the 
high-temperature entropy curves of different spin-$S$ cases.

The results in Fig.~\ref{Fig:NCTO}\textbf{d} indicate that the spin 
fractionalization also occurs in higher-spin Kitaev systems, and also 
huge low-temperature entropies associated with topological excitations. 
Due to the larger spin quantum number $S$, there are larger entropies 
and thus cooling capacity in the spin-1 case than that of the spin-$1/2$ 
case. Based on the simulations, we expect the high-$S$ Kitaev materials 
may serve as excellent refrigerants, and also notice that there are recent 
progresses in Kitaev magnets with higher spin $S$, including the spin-$1$ 
compound Na$_3$Ni$_2$BiO$_6$~\cite{Shangguan2023} and spin-$3/2$ 
CrSiTe$_3$~\cite{XuPRL2020}.
}
\\

\noindent{\textbf{Derivation of the equation of state in KFL.}} 
At zero field, the $\pi$-fluxes are virtually non-interacting between 
the two temperature scales $T_{\rm{L}}$ and $T_{\rm{H}}$, giving rise to a 
paramagnetic behavior described by a concise equation of state. To derive 
the equation of state for the Kitaev paramagnetism \rev{in the intermediate 
temperature regime,} we start with the Hamiltonian
\begin{equation}
H=K \sum_{\langle i,j\rangle_{\gamma}} S_i^{\gamma}S_j^{\gamma}
- B\sum_{i,\gamma} S_i^{\gamma} \equiv H_0 + H',
\end{equation}
where $H_0$ and $H'$ are non-commutative, and $H'$ is a perturbation containing
three $S^\gamma$ components coupled to a small field $B$.
We consider the orthonormal basis labeled as $|E_{\{W_P\}}^n\rangle$ as $n$-th state 
with the flux configurations $\{W_P\}$, {and $|E_{\{W'_P\}}^{n'}\rangle$
represents a $n'$-th state in the flux-flipped sector $\{ W'_P \}$}. 
The operator $S_i^{\gamma}$ 
applied on a site $i$ can flip two adjacent $\pi$ fluxes with a shared 
$\gamma$ bond. 
{Exploiting the Kubo formula, the susceptibility can be expressed as
\begin{equation}
\chi = \sum_{j, \gamma} \int_{0}^{\beta}\langle S_{i_0}^{\gamma}(\tau) \, S_j^{\gamma}\rangle_{\beta}d\tau,
\label{EqDynaSus}
\end{equation}
where $S_{i_0}^{\gamma}(\tau) = e^{\tau H} S_{i_0}^{\gamma} e^{-\tau H}$,
and $j$ runs over nearest-neighbor sites of
$i_0$ by $\gamma$ bond (as well as $i_0$ itself) in the fractional liquid regime
due to the extremely short-range correlations.
By inserting the orthonormal basis, we obtain the Lehmann spectral representation
\begin{equation}
\begin{split}
\langle S_{i_0}^{\gamma}(\tau) S_j^{\gamma}\rangle_{\beta}  = &  \sum_{\{W_P\},n} \sum_{n'}  e^{- \beta E^n_{\{ W_P \}}} \, e^{- \tau
\Delta_{n, \{W_P\};n',\{W'_P\}} } \, \\
& \langle E^n_{\{W_P\}} | S_{i_0}^{\gamma} | E^{n'}_{\{W'_P\}} \rangle
\langle E^{n'}_{\{W'_P\}} | S_{j}^{\gamma} | E^{n}_{\{W_P\}} \rangle,
\end{split}
\end{equation}
}
As the Majorana fermions are only weakly coupled to the $Z_2$ flux in the 
intermediate-temperature regime~\cite{Nasu2015}, $\Delta$ mainly represents 
the flux excitation gap,
{i.e., $\Delta_{n,\{W_P\};n',\{W'_P\}} \simeq
(E^{n'}_{\{W'_P\}}-E^n_{\{ W_P \}}) \sim T_L \ll T \equiv 1/\beta$.
Therefore, the decay factor $e^{- \tau \Delta_{n,\{W_P\};n',\{W'_P\}}} \simeq 1$,
thus $\langle S_{i_0}^{\gamma}(\tau) S_j^{\gamma}\rangle_{\beta}$
is virtually $\tau$-independent and $\chi$ can be expressed as
$\chi \simeq \frac{1}{T} \sum_{j,\gamma} \langle S_{i_0}^{\gamma} S_j^{\gamma} \rangle_\beta$
in the KFL regime.}
{As $C_{\rm K} \equiv \sum_{j,\gamma}\langle S_{i_0}^{\gamma}S_j^{\gamma} 
\rangle_{\beta}$ is nearly a constant below $T_{\rm H}$ (see Supplementary Note \B{1}),}
the susceptibility is therefore
\begin{equation}
\chi \approx \frac{C_{\rm K}}{T},
\end{equation}
and the equation of state for KFL is
\begin{equation}
M \approx \frac{C_{\rm K} B}{T}.
\end{equation}

{Using the Maxwell relation $(\partial M / \partial T)_{B} = (\partial S_{\rm m}/ \partial B)_T$, 
we express the magnetic entropy as $S_{\rm m} =  -\frac{C_{\rm K} B^2}{2T^2} + S_0(T)$.}
Therefore, $S_{\rm \pi-flux} \approx \frac{1}{2}\ln 2 - \frac{C_{\rm K}B^2}{2T^2}$ 
represents the $\pi$-flux part in the intermediate-temperature regime,
and the isentropes are mainly determined by $S_{\rm \pi-flux}$,
which constitute a series of lines through the origin, i.e.,
\begin{equation}
\frac{T}{B} = {\rm const}.
\end{equation}

\noindent{\bf{Data availability}}\\
Source data are provided with this paper.
{The data generated in this study have been deposited in the Zenodo database 
[https://doi.org/10.5281/zenodo.12736810].}\\

\noindent{\bf{Code availability}}\\
All numerical codes in this paper are available 
upon request to the authors. \\

%  ====== Bib ======= %
\bibliography{kitaevRef}

$\,$\\
\textbf{Acknowledgements} \\
H.L. and W.L. would like to thank Yuan Li and Xi Lin for stimulating discussions. 
The authors acknowledge supports by the National Natural Science Foundation 
of China (Grant Nos. 12222412 and 12047503), Strategic Priority Research 
Program of CAS (Grant No. XDB28000000), 
CAS Project for Young Scientists in 
Basic Research (Grant No.~YSBR-057), and China National Postdoctoral 
Program for Innovative Talents (Grant No. BX20220291). We thank HPC-ITP 
for the technical support and generous allocation of CPU time.

$\,$\\
\textbf{Author contributions} \\
H.L. and W.L. initiated this work. H.L., N.X., and Y.G. performed the tensor-network 
calculations. H.L., E.L., Y.Q., W.L., and G.S. analyzed the data and conducted 
theoretical analysis. H.L., G.S., and W.L. prepared the manuscript with input 
from all authors.

$\,$\\
\textbf{Competing interests} \\
The authors declare no competing interests. 

$\,$\\
\textbf{Additional information} \\
\textbf{Supplementary Information} is available in the online version of the paper. \\
\noindent

\clearpage
%======================
\newpage
\onecolumngrid
\begin{center}
{\large Supplementary Information for}
$\,$\\
\textbf{\large{Magnetocaloric Effect of Topological Excitations in Kitaev Magnets}}

$\,$\\
Li \textit{et al.}
\end{center}

\newpage

%=====================

\date{\today}

\setcounter{subsection}{0}
\setcounter{figure}{0}
\setcounter{equation}{0}
\setcounter{table}{0}

\renewcommand{\thesubsection}{\normalsize{Supplementary Note \arabic{subsection}}}
\renewcommand{\theequation}{S\arabic{equation}}
\renewcommand{\thefigure}{\arabic{figure}}
\renewcommand{\thetable}{\arabic{table}}

\subsection{Thermal fractionalization in the Kitaev spin liquid}
\label{SecSM:Frac}

To reveal the spin fractionalization in the Kitaev model under magnetic fields, 
in Supplementary Fig.~\ref{Fig:SMfractionalization} we show XTRG results 
on a YC$4\times 10\times 2$ lattice, and apply small magnetic fields along [1 1 1] 
direction. In Supplementary Figs.~\ref{Fig:SMfractionalization}\textbf{a,d}, 
the expectation values of the plaquette operator, also dubbed localized $Z_2$ 
gauge flux, are defined as 
\begin{equation}
{\langle W_{\rm P} \rangle} = \langle \sigma_i^x \sigma_j^y \sigma_k^z \sigma_l^x 
\sigma_m^y \sigma_n^z 
\rangle,
\end{equation}
where the set $\{i,j,k,l,m,n\}$ are the six vertices in a hexagon plaquette. The plaquette 
operator $W_{\rm P}$ is a conserved quantity in the pure Kitaev model~\cite{Kitaev2003,
Kitaev2006}, {whose expectation value $\langle W_{\rm P} \rangle$} rapidly increases 
at $T\simeq T_{\rm L}$ {as shown in Supplementary Fig.~\ref{Fig:SMfractionalization}\textbf{a},} and finally converges to $\langle 
W_{\rm P} \rangle = 1$ in the low-temperature limit. Under finite fields, $W_{\rm P}$ no 
longer commutes with the Hamiltonian, yet it still show similar behaviors as in 
the pure Kitaev case, except for the smaller converged value $\langle W_{\rm P} 
\rangle < 1$ below $T_{\rm L}$~\cite{Li2020b}.

% ================== spin fractionalization ================== %
\begin{figure}[h!]
\includegraphics[angle=0,width=1\linewidth]{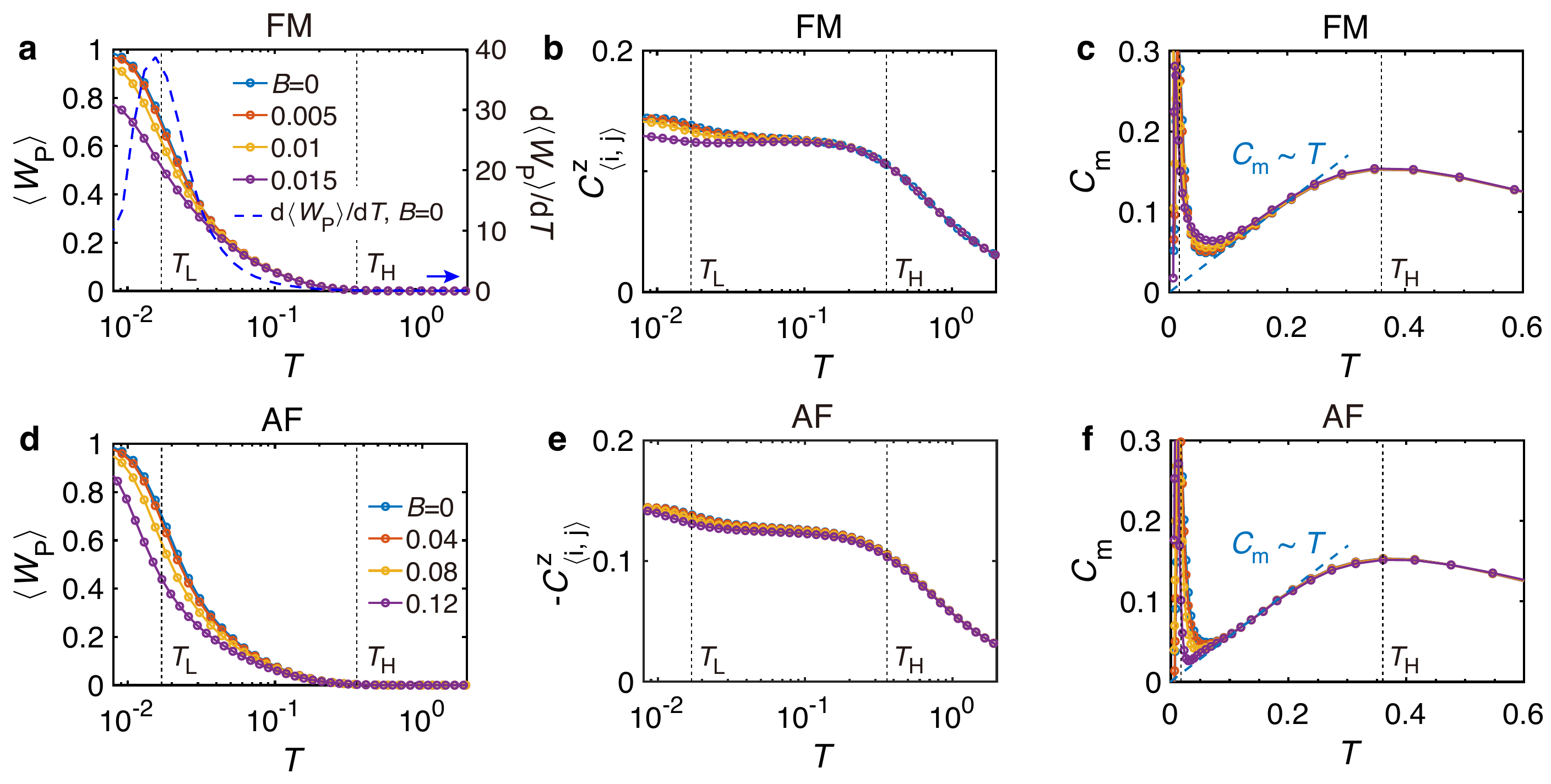}
\renewcommand{\figurename}{\textbf{Supplementary Figure }}
\caption{Various thermodynamic properties, including \textbf{a,d} the expectation
values of $Z_2$ flux $\langle W_{\rm p} \rangle$, \textbf{b,e} the $z$-component 
of spin correlations $C_{\langle i,j \rangle}^{z}$ measured on the nearest $z$-type 
bond with field-induced background subtracted, and \textbf{c,f} specific heat 
$C_{\rm m}$ curves of \textbf{a,b,c} ferromagnetic (FM) and \textbf{d,e,f} antiferromagnetic (AF) Kitaev models under 
small fields. Two temperature scales $T_{\rm H}$ and $T_{\rm L}$ for $B=0$ are 
marked by the vertical dashed lines in all panels. {In panel \textbf{a}, the derivatives 
${\rm d}\langle W_P \rangle/{\rm d}T$ at zero field are shown with the blue dashed curve.} 
In panels \textbf{c,f}, the metallic behavior of $C_{\rm m}$ at intermediate-temperature 
KFL regime is indicated by the blue dashed line.
}
\label{Fig:SMfractionalization}
\end{figure}
% ================================================== %

The bond-dependent short-range spin correlation on the nearest $\gamma$ 
bond is defined by
\begin{equation}
{C}^{\gamma}_{\langle i,j\rangle}=[\langle S_i^{\gamma} S_j^{\gamma} \rangle 
- \langle S_i^{\gamma}\rangle \langle S_j^{\gamma} \rangle],
\end{equation}
and we show the results of $\gamma=z$ bond in Supplementary 
Figs.~\ref{Fig:SMfractionalization}\textbf{b,e}. 
{Under zero field, the $z$-component} spin correlations rapidly establish 
at the high temperature scale $T\simeq T_{\rm H}$, and remain almost the 
same value down to low temperatures (except for the slight upturn at around
$T_{\rm L}$). {The $x$- and $y$-component spin correlations remain zero 
across the entire temperature regime, due to bond-oriented spin correlations 
in the Kitaev model.} As we apply a small $B$ field, the low-temperature scale 
$T_{\rm L}$ moves towards lower temperature, while $T_{\rm H}$ remains 
unchanged, resulting in a broader intermediate-temperature regime.

As the magnetic entropy releases half of its total value near each of the two 
temperature scales $T_{\rm H}$ and $T_{\rm L}$~\cite{Nasu2015,Li2020b}, 
we dub the intermediate regime with fractional  ($\sim \frac{1}{2} \ln{2}$) entropy
as Kitaev fractional liquid (KFL). In the KFL regime, magnetic specific heat shows 
a linear-$T$ scaling, i.e., a metallic behavior, as indicated by the dashed 
line $C_{\rm m} \sim T$ in Supplementary Figs.~\ref{Fig:SMfractionalization}\textbf{c,f}.

% =========== Out-of-plane fields ======== %
\subsection{Additional results of the AF Kitaev model under magnetic fields}
\label{SecSM:Convergence}

\textbf{Data convergence.}
In Supplementary Fig.~\ref{Fig:Convergence}, we show the convergence
of the AF Kitaev model with $B=0.3$ in the intermediate-field regime. As we 
increase the retained bond dimension $D$, the specific heat [c.f., Supplementary 
Fig.~\ref{Fig:Convergence}\textbf{a}] and the thermal entropy [c.f., Supplementary 
Fig.~\ref{Fig:Convergence}\textbf{b}] results exhibit only small changes, indicating 
that the calculations on the YC4$\times$10$\times$2 system has been converged 
vs. bond dimension. Moreover, in Supplementary Fig.~\ref{Fig:Convergence}\textbf{c} 
we compare the entropy curves on YC4$\times$10$\times$2 and 
XC8$\times$5$\times$2 lattices, and find the finite-size effects are also small, 
as evidenced by the robustness of the low-temperature scale $T^*_{\rm L}$ 
on different, YC and XC, cylindrical geometries.

% ================== convergence ================== %
\begin{figure}[h!]
\includegraphics[angle=0,width=0.99\linewidth]{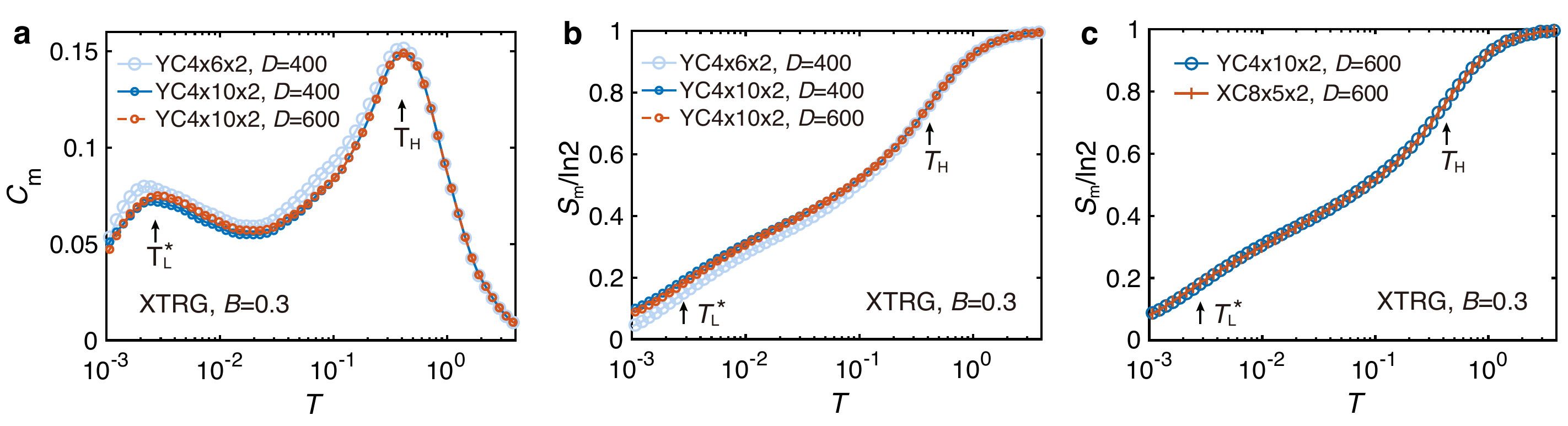}
\renewcommand{\figurename}{\textbf{Supplementary Figure }}
\caption{\textbf{a} The specific heat $C_{\rm m}$ and \textbf{b} thermal entropy 
$S_{\rm m}/\ln{2}$ curves of AF Kitaev model on various YC geometries with 
intermediate field $B=0.3$. The temperature scales $T_{\rm H}$ and 
$T^*_{\rm L}$ are indicated by the arrows.
\textbf{c} The entropy curves of YC4$\times$10$\times$2 and 
XC8$\times$5$\times$2 lattices under field of $B=0.3$, 
where ``XC'' means the periodic boundary condition (BC) 
along the $L=5$ (zigzag) direction and open BC along the 
width $W=8$ direction.}
\label{Fig:Convergence}
\end{figure}
% ================================================== %

\textbf{Correlation length.}
In order to check the finite-size effects in the thermodynamic quantities, 
in particular the properties down to the low-temperature scale $T^*_{\rm L}$ 
in the intermediate-field phase, we perform calculations of the correlation length 
$\xi$ with $B=0.3$ at various temperatures on the YC4$\times$10$\times$2 lattice.

As shown in Supplementary Fig.~\ref{Fig:xi}, the spin correlations 
$C_{\rm ij} = \langle {\textbf S}_i \cdot {\textbf S}_j \rangle 
- \langle {\textbf S}_i\rangle  \cdot \langle {\textbf S}_j \rangle$
versus real-space distance $R_{\rm ij}$
between sites $i$ and $j$ are plotted in a semi-logarithmic scale,
where $i$ denotes the central site and $j$ runs along the zigzag chain.
As the system cools down, the fitted value of the correlation length 
$\xi$ exhibits an increasing trend with decreasing temperature,
reaching about $0.71$ at $T\simeq0.001$. As it is even much smaller 
than the finite width ($W=4$) of the cylinder adopted in our simulations,
we can therefore conclude that the low-temperature results are less 
affected by finite-size effects. It is noteworthy that, as compared with 
the ground-state data extracted from Ref.~\cite{Patel2019}, we find 
the DMRG data actually shows much longer correlation length, offering 
another reason to do finite-temperature simulations on cylinder geometries. 
Therefore, we believe that the specific heat and entropy curves 
shown in the main text, including data near the low-temperature scale 
$T^*_{\rm L}$, reflect intrinsic 
properties not much influenced by finite-size effects in the system.

% ================== correlations ================== %
\begin{figure}[h!]
\includegraphics[angle=0,width=0.6\linewidth]{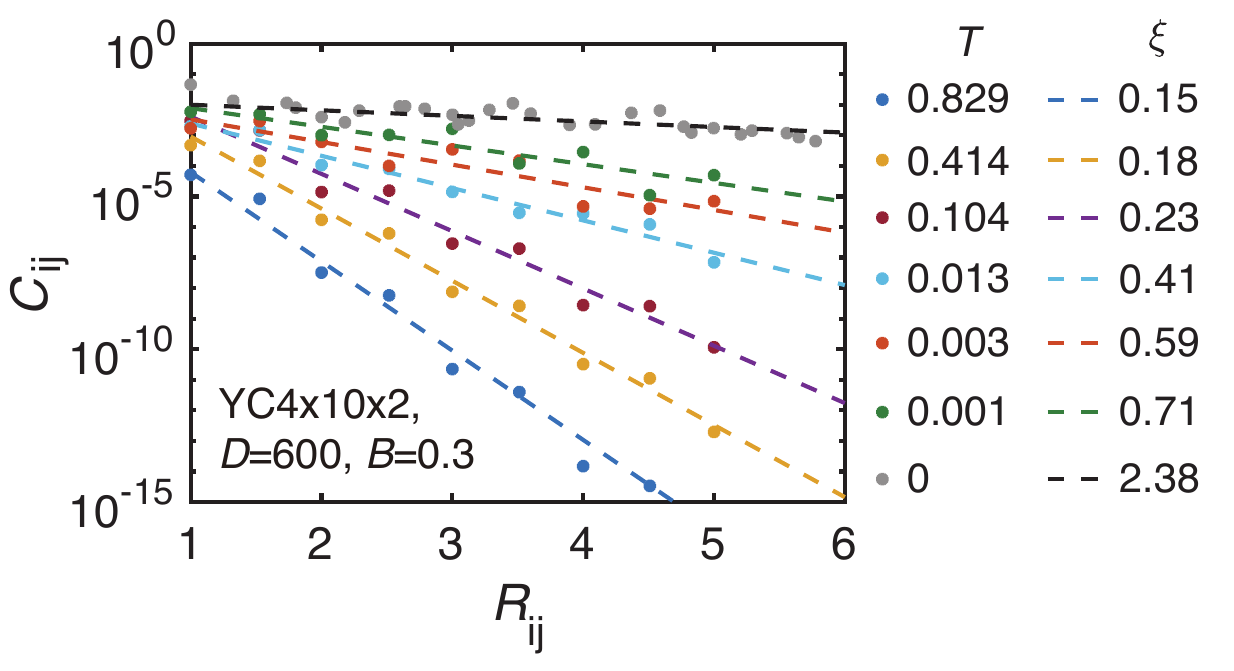}
\renewcommand{\figurename}{\textbf{Supplementary Figure }}
\caption{The spin correlations $C_{\rm ij}$ of AF Kitaev model on the
YC4$\times$10$\times$2 lattice under intermediate field $B=0.3$ 
at various temperatures. The exponential fitting curves $e^{-R_{\rm ij}/\xi}$ 
are shown with dashed lines correspondingly.
}
\label{Fig:xi}
\end{figure}
% ================================================== %

\textbf{Matrix product operator entanglement.}
In XTRG, the thermal density matrix $\rho(\beta = 2^n\tau_0)$ has been 
expressed in the form of matrix product operator (MPO), thus we can compute 
the bipartite entanglement entropy $S_{\rm E}(\beta)$ from the purified 
supervector $\ket{\rho(\beta/2)}$.
For example, $S_{\rm E}$ alway saturates or peaks at $T\lesssim \Delta$ 
with $\Delta$ the excitation gap for a gapped system. On the contrary, 
in the gapless phase it may diverge as temperature lowers.

% ================== MPO entanglement ================== %
\begin{figure}[h!]
\includegraphics[angle=0,width=0.8\linewidth]{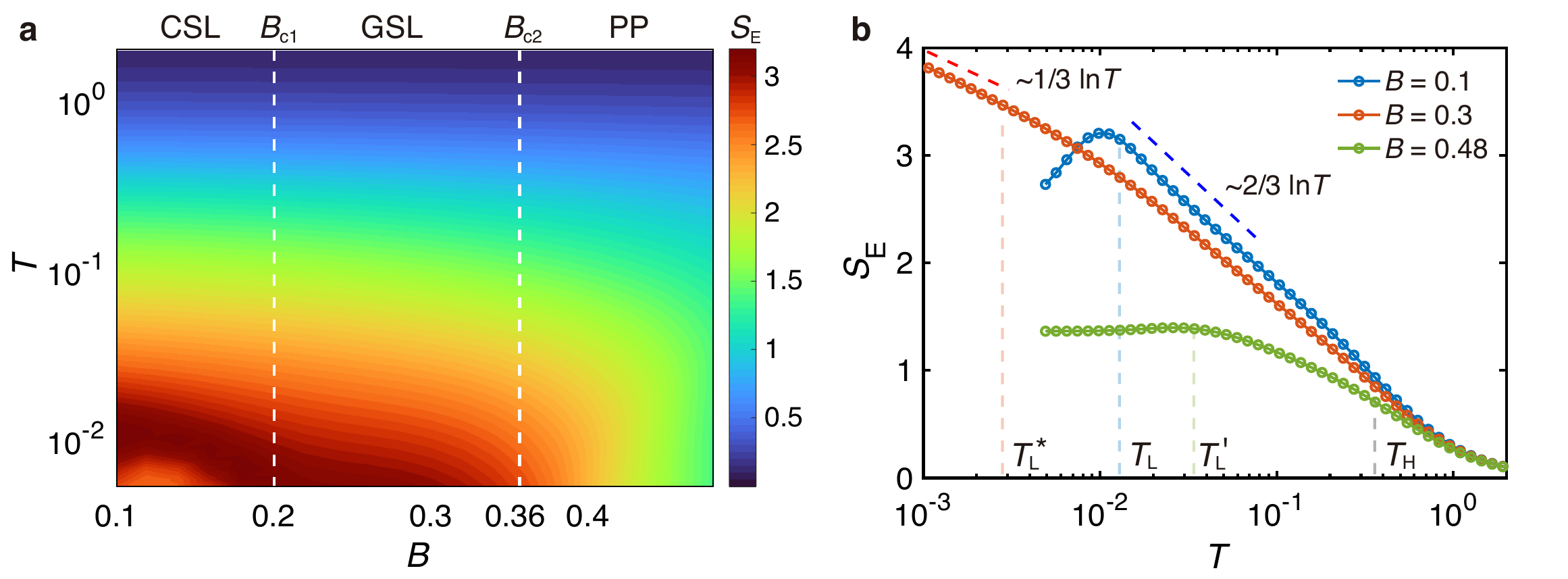}
\renewcommand{\figurename}{\textbf{Supplementary Figure }}
\caption{\textbf{a} The contour plot of MPO entanglement $S_{\rm E}$ of 
AF Kitaev model in magnetic fields. The white dashed lines indicate the 
two quantum phase transitions {at $B_{c1}$ and $B_{c2}$} determined 
from ground-state DMRG calculations. Various temperature-dependent 
$S_{\rm E}$ curves are shown in panel \textbf{b}, where the logarithmic 
fittings for the small- and intermediate-field regimes are marked by blue 
and red dashed lines, repsectively.
}
\label{Fig:SE}
\end{figure}
% ================================================== %

In Supplementary Fig.~\ref{Fig:SE}\textbf{a}, we show the landscape of bipartite 
entanglement entropy $S_{\rm E}$ under fields $B$ applied along [111] direction. 
At small fields, i.e., $B<0.2$, $S_{\rm E}$ firstly increase and then drops as 
temperature lowers, consistent with the fact that the chiral spin liquid phase
has a very small vison gap comparable to the lower temperature scale $T_{\rm L}$. 
At intermediate fields, i.e., $0.2<h<0.36$, the $S_{\rm E}$ value continues to 
increase at low temperature, supporting a gapless nature of intermediate quantum 
spin liquid phase. At field $B>0.36$, $S_{\rm E}$ converges to a small finite value 
in the low-temperature limit, confirming the gapped nature of the (partially) polarized states. 

In Supplementary Fig.~\ref{Fig:SE}\textbf{b}, we show typical $S_{\rm E}$ curves. 
For the $B=0.1$ case, a peak can be clearly observed near the low-temperature 
scale $T_{\rm L}$, above which the data exhibits a logarithmic scaling as $S_{\rm E} 
\sim -2/3$ ln$T$ in the KFL regime. This scaling is robust and we stress that a similar 
$-2/3 \ln{T}$ scaling has been observed in the ferromagnetic Kitaev model under [001] 
field along a different direction~\cite{Li2020b}. In contrast, for $B=0.3$ case the $S_{\rm E}$ 
diverges logarithmically till the lowest temperatures, fitted with a scaling $S_{\rm E} \sim -1/3 
\ln{T}$ as indicated by the red dashed line. 
The exponent $1/3$ suggests that to simulate the width-4 Kitaev cylinder system in the 
intermediate-field phase, the required computational resource scales similarly to that 
near a conformal quantum critical point with central charge $c=1$. In addition, the maximal 
$S_{\rm E}$ value is less than 4 at temperature down to $T = 0.001$, which indicates 
that the YC4$\times$10$\times$2 systems can be well simulated without requiring an 
excessively large number of bond states. In addition, the high-field curves, e.g., the results 
with $B=0.48$, saturate at a relatively high temperature scale $T'_{\rm L}$ as expected.

% ================== MH and Gruneisen ================== %
\begin{figure}[h!]
\includegraphics[angle=0,width=0.5\linewidth]{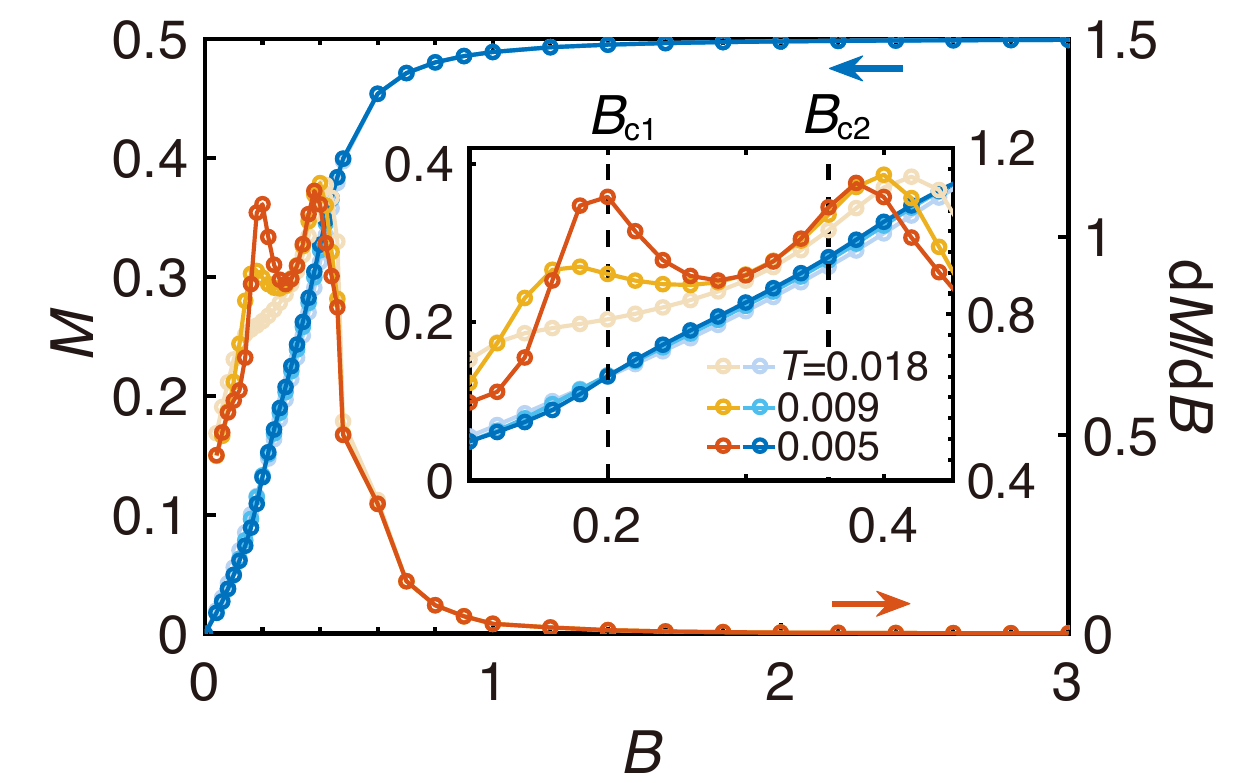}
\renewcommand{\figurename}{\textbf{Supplementary Figure }}
\caption{The magnetization curves $M$ vs. $B$ and its derivatives 
${\rm d}M/{\rm d}B$ at different temperatures. The two transition 
fields $B_{\rm c1}$ and $B_{\rm c2}$ are identified by the peaks 
of ${\rm d}M/{\rm d}B$, and are marked by two black dashed lines 
in the inset.}
\label{Fig:MH}
\end{figure}
% ================================================== %

\subsection{Field-induced  quantum phase transitions in the AF Kitaev model}
\label{Sec: MHcurves}

\textbf{Low-temperature magnetization curves.}
In this section,
we show the magnetization $M$ curves and their derivatives ${\rm d}M/{\rm d}B$
at various low temperatures for the AF Kitaev model, to confirm the consistency 
of our low-temperature XTRG data and the ground-state DMRG calculations
on determining the transition fields $B_{\rm c1}$ and $B_{\rm c2}$.

In Supplementary Fig.~\ref{Fig:MH}, one can see that the magnetic moment 
at low temperatures grows with $B$. At $B\simeq 0.6$, the magnetization reaches 
90~\% of its saturation value, in agreement with the previous ground-state simulations~\cite{Gohlke2018}. Moreover, at low temperature there are 
double peaks in the derivatives ${\rm d}M/{\rm d}B$, which correspond 
to the two transition fields. In the inset, we zoom in both curves, and find
the peaks of ${\rm d}M/{\rm d}B$ converge to the two quantum phase 
transition fields $B_{\rm c1}$ and $B_{\rm c2}$ determined from the
ground-state results~\cite{Zhu2018,Jiang2019,Patel2019}, as indicated by 
vertical black dashed lines in Supplementary Fig.~\ref{Fig:MH}.

% ================== structure factor ================== %
\begin{figure}[h!]
\includegraphics[angle=0,width=0.85\linewidth]{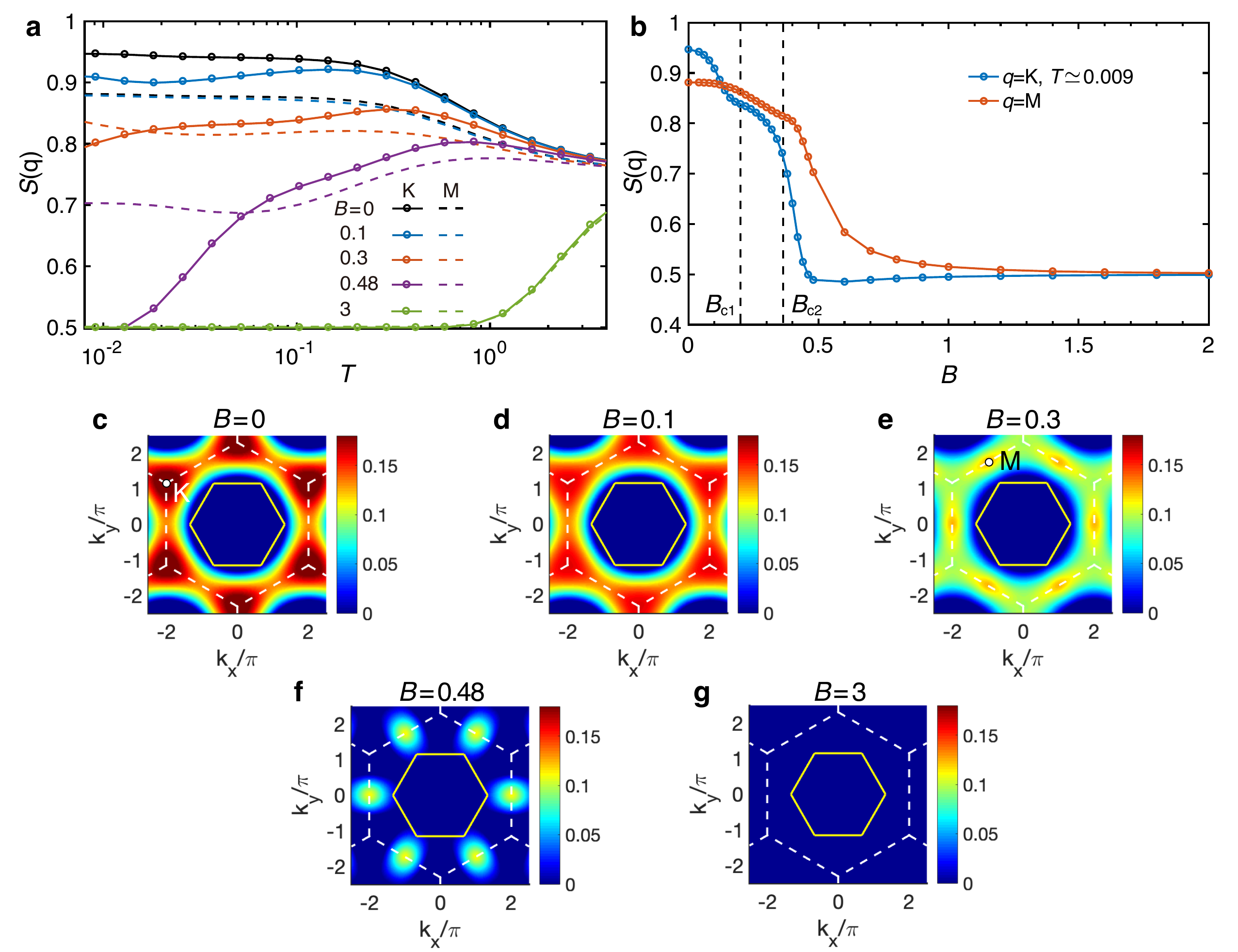}
\renewcommand{\figurename}{\textbf{Supplementary Figure }}
\caption{\textbf{a} The temperature-dependent spin-structure factors $S(q)$ 
at $q$=K and M points computed in the AF Kitaev model under various fields. \textbf{b} 
The field-dependent $S(q)$ curves at a low temperature $T\simeq0.009$, where the 
black dashed lines indicate the ground state transition fields $B_{\rm c1}$ and 
$B_{\rm c2}$~\cite{Zhu2018,Jiang2019,Patel2019}. \textbf{c-g} The landscapes of $S(q)$ 
with various $B$ measured at a low temperature of $T\simeq0.009$. The round 
peak firstly appears at the K point, then moves to the M point as field increases, 
and finally disappears at large fields.
}
\label{Fig:Sq2}
\end{figure}
% ================================================== %

\textbf{Spin structure factors.}
We calculate the temperature- and field-dependent spin-structure factors 
${S}(\textbf{q})=\sum_{j\in {N}} 
e^{ i \textbf{q} (\textbf{r}_j-\textbf{r}_{i_0})} 
(\langle S_{i_0} S_j\rangle - \langle 
S_{i_0}\rangle\langle S_{j}\rangle)$,
where $i_0$ represents a central reference site, and the field-induced 
uniform background has been subtracted. We simulate the structure 
factor $S({\textbf q})$ to identify the phase boundary and the nature of 
the low-temperature phases in the AF Kitaev model. 

In Supplementary Fig.~\ref{Fig:Sq2}\textbf{a}, we show the $S({\rm K})$ 
and $S({\rm M})$ curves at various fields. At small fields, e.g., $B=0$ and 
$0.1$, the values of $S({\rm K})$ are always larger than the $S({\rm M})$ 
values as temperature decreases, and the structure factors thus show peaks 
at K points [c.f., Supplementary Figs.~\ref{Fig:Sq2}\textbf{c,d}] in the CSL 
phase. When field further increases, the M-point peaks become greater 
than the K-point ones at low temperatures, which can be seen in, e.g., 
Supplementary Fig.~\ref{Fig:Sq2}\textbf{e}. In the high-field limit, both 
the $S({\rm K})$ and $S({\rm M})$ peaks are suppressed [c.f., Supplementary 
Fig.~\ref{Fig:Sq2}\textbf{g}].

The quantum phase transitions could also be identified 
from the low-temperature structure factors $S({\textbf q})$.
In Supplementary Fig.~\ref{Fig:Sq2}\textbf{b}, we show the $S({\textbf q})$ 
curves versus fields calculated at a low temperature $T\simeq 0.009$.
The transition fields are marked by the two black dashed 
lines with $B_{\rm c1}$ and $B_{\rm c2}$. As illustrated in Supplementary
Figs.~\ref{Fig:Sq2}\textbf{c-g}, we find for $B < B_{c1}$ the $S(\rm K)$ is 
more pronounced, which exhibits a sudden decrease at around $B_{\rm c1}$ 
where M-point peak exceeds that of the K point. In the intermediate phase, 
the peak strengths of $S(\rm K)$ and $S(\rm M)$ are comparable [with M-point 
sightly higher, as evidenced by the bright blobs in Supplementary Fig.
\ref{Fig:Sq2}\textbf{e}]. For $B> B_{\rm c2}$ both intensities show steep 
drop as the systems enters the partially polarized phase.

{\subsection{The cooling efficacy of FM and AF Kitaev magnets}}
\label{Sec:cooling}

{Any practical paramagnetic coolant would have a cut-off refrigeration temperature, 
usually in the same order of residual spin-spin coupling $J$ in the system. Here for
the FM Kitaev system, the low cut-off temperature $T_{\rm L}$ is about two orders of 
magnitude lower than the spin-spin coupling $|K|$, which is very remarkable and can
give rise to very low cooling temperature. On the other hand, for the AF Kitaev case 
there emerges a gapless U(1) QSL in the intermediate fields and thus the cut-off temperature 
is absent, rendering it also a very appealing refrigerant.}

{To illustrate the magnitude of the cooling temperatures,  
we translate the natural units into those commonly used in experiments,
by assigning a value of $|K| = 1.34$ Kelvin and a Landé g-factor of $g=2$. 
It follows that a magnetic field of $B/|K|=1$ corresponds to a strength of 1 
Tesla. Under these conditions, the efficacy of the FM Kitaev material in 
cooling is evident in Supplementary Figs.~\ref{Fig:S7}\textbf{a,b}. 
Starting from 0.5 K and initial field of 0.8 T, the lowest temperature achieved 
through adiabatic demagnetization is about 13 mK. In Supplementary Fig.~\ref{Fig:S7}\textbf{b}, 
when the initial condition is changed to 1~K and 0.8~T, 
the lowest temperature is 35~mK. On the other hand, the calculated results for 
AF Kitaev model are shown in Supplementary Fig.~\ref{Fig:S7}\textbf{c}, 
where the lowest cooling temperature would be below 10~mK through an 
adiabatic process, with field decreasing from $B_{\rm i} = 3$~T to $B_{\rm c1} \simeq 0.2$~T.}

%=========== Fig.R5, ===========%
\begin{figure}[h!]
\includegraphics[angle=0,width=0.98\linewidth]{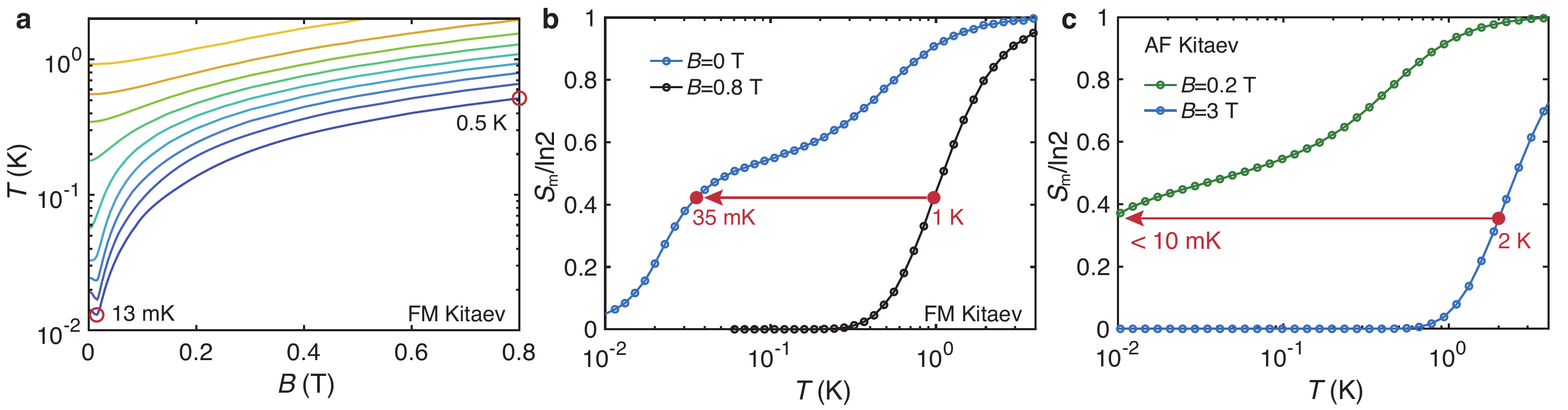}
\renewcommand{\figurename}{\textbf{Supplementary Figure }}
\caption{{{\textbf{a} Isentropes and \textbf{b} entropy curves of the FM Kitaev 
model. \textbf{c} The entropy curves of the AF Kitaev model. In units of Kelvin 
and Tesla, we show that sub-Kelvin and even millikelvin low temperature can be 
achieved for both FM and AF Kitaev systems with $|K|=1.34$~Kelvin, driven by 
a moderate magnetic field change.}}}
\label{Fig:S7}
\end{figure}
% =============================================== %\

{The above results offer a glimpse into the remarkable potential of Kitaev materials 
as ultra-efficient coolants, underlining their great potential in achieving low temperatures 
necessary for cutting-edge applications such as quantum technologies. Potential candidate 
materials, with a moderate interaction strength, i.e., on the order of Kelvins, include 
BaCo$_2$(AsO$_4$)$_2$~\cite{Zhang2023NatMat,Tu2022arXiv,Zhong2020SA}, 
as well as some rare-earth chalcohalides such as Ba$_9$Yb$_2$(SiO$_4$)$_6$
\cite{Tian2023Honeycomb}. }

\end{document}